\documentclass[12pt]{article}

\usepackage[margin=1in]{geometry}
\usepackage[T1]{fontenc}
\usepackage[utf8]{inputenc}

\usepackage{amsmath,amssymb}

\usepackage{graphicx}
\usepackage{booktabs}
\usepackage{multirow}
\usepackage{float}
\usepackage{subcaption}   
\usepackage{caption}
\usepackage{mdframed}
\usepackage{makecell}

\usepackage{multicol}

\usepackage{colortbl}
\usepackage[table,xcdraw]{xcolor}

\usepackage{arydshln}

\usepackage{tikz}
\usetikzlibrary{positioning,fit,backgrounds}

\usepackage[hidelinks]{hyperref}
\usepackage{xurl}

\usepackage{pdfpages}

\title{Multi-Ecosystem Modeling of OSS Project Sustainability}

\author{
  Arjun Ashok$^{*}$,\;
  Nafiz Imtiaz Khan$^{*}$,\;
  Swati Singhvi,\; \\
  Stefan Stanciulescu,\;
  Zhouhao Wang,\;
  Vladimir Filkov \\[4pt]
  Department of Computer Science, University of California, Davis \\
  Davis, CA 95616, USA \\[2pt]
  {\small $^{*}$Both authors contributed equally to this work.}
}

\date{}  

\begin{document}
\maketitle

\begin{abstract}
OSS projects join software foundations like Apache, Eclipse, and OSGeo to get
help with their immediate development plans and to improve long-term
sustainability prospects. Once inside, they receive governance advice,
incubation support, and community-building mechanisms. But foundations differ
in their policies, funding models, and support strategies, and thus, outcomes
can vary across them. Moreover, since projects have diverse needs and join
these foundations at different lifecycle stages, it can be challenging to
decide on the appropriate project-foundation match and on the project-specific
plan for sustainability.
Here, we present an empirical study and quantitative modeling analysis of the
sustainability of incubator projects in the Apache, Eclipse, and OSGeo
foundations, and, additionally, of GitHub projects from outside of the
foundations. We develop sustainability models based on projects' sociotechnical
traces and demonstrate their effectiveness within and across the foundations.
We show that our best model, which uses a project-foundation router, OSS-ProF,
can effectively forecast sustainability outcomes not only within but across
foundations. In addition, the generalizability of the framework allows us to
effectively apply the approach to GitHub projects outside the foundations. Our
study highlights the value of sociotechnical networks as a generalizable
framework for characterizing and addressing software project sustainability
issues.
\end{abstract}

\textbf{Keywords:} Open Source Software, Sustainability, Sociotechnical
Networks, Deep Neural Networks, Actionable Recommendations

\section{Introduction}

Modern social coding platforms and free-to-use software development tools make starting an open-source software (OSS) project easier than ever before. However, competition for programmers/funding and the complex nature of distributed OSS project communities make it challenging to sustain projects, as evidenced by the high failure rates of nascent projects \cite{41_schweik2012internet}.

To enhance their chances of becoming sustainable, projects join foundations, e.g., Apache Software Foundation (ASF) \cite{62_apache}, Eclipse Foundation (EF) \cite{eclipse}, OSGeo Foundation (OF) \cite{osgeo}, Linux Foundation (LF) \cite{linuxfoundation}, and others. Often, those foundations have incubators that implicitly or explicitly modify project activities and governance via tailored suggestions and support to the developer communities. However, foundations differ significantly in their policies, incubation models, funding, and project support mechanisms. Moreover, projects joining these foundations are also diverse, coming at different lifecycle stages and having different needs. This multivariate diversity makes it challenging to plan for project sustainability. 
Fortunately, most software foundations provide public access activity traces, which can be used to model sustainability.

Research on OSS sustainability has been on the rise. Yin et al.~\cite{2_yin2021sustainability} used sociotechnical features and AI methods (LSTMs) to model and predict the graduation or retirement of more than 200 ASF Incubator OSS projects. 
Parl et al.~\cite{43_park2024survivability} employed a polynomial regression model to predict the survivability of OSS through bug-fixing activities. The study used the 24 most recent projects available in the iOS version of the Kakaotalk platform \cite{61_kakao}. 
Xiao et al.~\cite{4_xiao2023early} predicted sustainability in the formative stages of GitHub OSS projects based on features such as the number of issues, pull requests, starred projects, etc. 
These studies provide valuable information on sustainability within individual platforms \cite{6_joblin2022successful}, yet leave open questions regarding generalizability. First, to the best of our knowledge, no prior work has examined the extent to which sustainability models vary across foundations. Second, the conceptual distinction between project success, used on GitHub, and sustainability, used in foundations, has not been empirically studied. Furthermore, the activities and governance structures of foundations such as ASF, EF, and OF vary widely. As a result, features predictive of sustainability in one foundation may be less relevant—or absent—in another.
Thus, to reason across foundations, we need powerful abstract modeling frameworks that are also generalizable.

Sociotechnical networks have shown great promise in studies of complex software development phenomena. They have been widely used to organize digital trace data~\cite{101_yin2022open,106_howison2011validity} and to model the structure~\cite{104_appelbaum1997socio,105_hong2017creating} and functioning~\cite{102_sarker2019socio, 103_bird2009putting} of software teams. Prior OSS sustainability studies have already demonstrated their effectiveness as predictive structures~\cite{2_yin2021sustainability, 100_ramchandran2022exploring}.
We also use them here, as a framework for characterizing projects.

Building on all of the above, we hypothesize that:

\noindent
\fbox{%
  \parbox{\linewidth}{%
    OSS project sociotechnical network profiles can be used to effectively predict project sustainability within and outside of foundation ecosystems.
  }%
}


To test this hypothesis, here we present an empirical study and quantitative analysis of the sustainability of incubator projects in the ASF, EF, and OF foundations, and, additionally, of OSS projects from GitHub (outside of foundations). Staring from projects' sociotechnical trace profiles, we develop foundation-specific sustainability forecasting models and demonstrate their effectiveness within foundations. We also developed a project-foundation classifier, and demonstrate its effectiveness on predicting project sustainability across and outside of the foundations. We find that:

\begin{enumerate}
\item Sustainability models trained on sociotechnical features are highly effective when applied within the same foundation;

\item Models exhibit limited transferability across foundations. While some general patterns exist, sustainability signals learned in one foundation do not fully translate to another foundation/ecosystem, showcasing the limitations of sociotechnical features in  creating a unified sustainability model;

\item Models trained on sustainability labels can reasonably predict success in GitHub projects, i.e., outside of a foundation; however, models trained on GitHub success fail to predict foundation-defined sustainability outcomes;

\item Sociotechnical features can be leveraged to train a project-foundation classifier, capable of predicting the most suitable software foundation model for each OSS project;

\item While there is feature overlap, the most predictive features vary across foundations, indicating that while some sociotechnical indicators are universal, others are ecosystem-specific. This reflects the diversity across foundations.


\end{enumerate}

To our knowledge, this is the first study to examine sustainability prediction in a multi-foundation setting. Our results demonstrate that sociotechnical networks can provide a robust and generalizable modeling framework, enabling broader study of sustainability.

\section{Background \& Theory}

\subsection{Sociotechnical Networks}
OSS project activities fall into two categories: technical (code commits, pull request reviews, issue resolutions) and social (email discussions, issue comments, community forums) \cite{32_wu2007investigating}. These can be captured as networks where nodes represent developers and files (technical) or people (social), with edges representing interactions \cite{33_wu2007analysis, 2_yin2021sustainability}.

Sociotechnical networks model the intricate relationships between social interactions and technical dependencies in software engineering environments \cite{107_scacchi2005socio,108_storey2020software}. They reveal structural interdependencies underlying effective coordination \cite{104_appelbaum1997socio,105_hong2017creating} and illuminate how collaboration impacts productivity, code quality, and project longevity \cite{101_yin2022open,106_howison2011validity}. Their ability to capture both relational dynamics and technical dependencies makes them particularly valuable for OSS sustainability research, where community engagement, contributor retention, and governance strategies are critical \cite{2_yin2021sustainability,100_ramchandran2022exploring}.

\subsection{Success and Sustainability in OSS Projects}
\label{sec:label_provenance}

Success and sustainability lack universally accepted measures across OSS foundations and platforms \cite{30_crowston2003defining}. Table~\ref{tab:label_provenance} formalizes the distinctions across ecosystems studied in this work.

\begin{table}[t]
\centering
\caption{Label Provenance Across Ecosystems}
\label{tab:label_provenance}
\small
\begin{tabular}{@{}p{1.8cm}p{1.5cm}p{3.5cm}p{2.2cm}p{1.8cm}p{1.5cm}@{}}
\toprule
\textbf{Ecosystem} & \textbf{Label} & \textbf{Definition} & \textbf{Decision Authority} & \textbf{Time Horizon} & \textbf{Sample Size} \\
\midrule
Apache (ASF) & Graduated / Retired & Self-sustaining community with governance, IP compliance, regular releases & PMC + Board vote & 1–3 years incubation & 262 (205/57) \\
\addlinespace
Eclipse (EF) & Graduated / Archived & Stable codebase, diverse contributors, Eclipse-compliant releases & PMC + EDP review & 1–2 years incubation & 161 (142/19) \\
\addlinespace
OSGeo (OF) & Graduated / Withdrawn & Active geospatial project with community engagement, governance & Incubation Committee & Variable (multi-tier) & 20 (13/7) \\
\addlinespace
GitHub & Successful / Unsuccessful & Sustained activity, contributor retention, community growth & Researcher-assigned (Joblin et al.) & Early development phase & 21 (5/16) \\
\bottomrule
\end{tabular}
\end{table}

Success typically reflects widespread adoption, active engagement, technical innovation, and measurable domain impact \cite{29_weber2004success}. Sustainability reflects a project's capacity to endure and remain relevant over time through stable governance, consistent contributor participation, resource continuity, and adaptability \cite{31_gamalielsson2014sustainability}. 

Within foundations, projects deemed sustainable are labeled "graduated," indicating self-sustaining communities, while "retired" projects are considered unsustainable due to limited activity, waning relevance, or insufficient engagement. Importantly, retirement can represent a natural lifecycle phase when a project's purpose is fulfilled or superseded, rather than outright failure. Both graduation and retirement maintain balanced, healthy OSS ecosystems.

\textbf{Construct Relationship.} We do not claim success $\equiv$ sustainability; rather, \textit{sustainability subsumes a subset of success signals}. GitHub success metrics (stars, forks, contributor activity) may indicate short-term momentum, while foundation sustainability requires demonstrated governance maturity, licensing compliance, community diversity, and release capability. Our cross-ecosystem modeling (RQ3) empirically examines this construct alignment rather than assuming label equivalence.

\subsection{OSS Foundations/Ecosystems}

\textbf{Apache Software Foundation Incubator (ASFI).}
\label{ASFI}
The ASFI \cite{62_apache} provides structured mentorship for projects joining the ASF ecosystem \cite{15_1apachefoundation}, guided by the "Apache Way" philosophy: "If it didn't happen on the mailing list, it didn't happen" \cite{23_apacheway, 2_yin2021sustainability}. 

\begin{figure}[!ht]
\includegraphics[width = \linewidth, height=5cm]{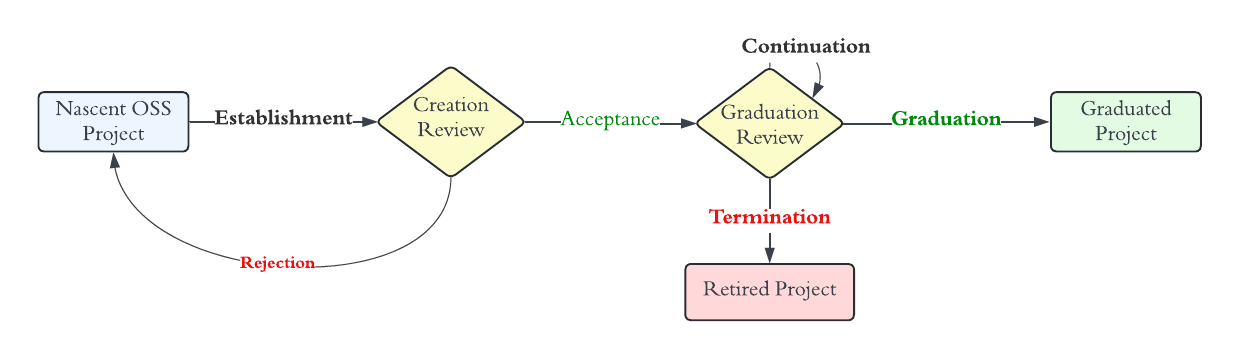}

\caption{Incubation Process of Apache Foundation}
\label{incubation_apache}
\end{figure}

Incubation typically spans one to three years \cite{15_2apachefoundation}. Projects must build diverse contributor communities, establish transparent governance, ensure licensing compliance, and demonstrate regular ASF-compliant releases \cite{15_1apachefoundation, 15_2apachefoundation}. Figure~\ref{incubation_apache} shows the multi-stage review pipeline from Creation Review through graduation, continuation, or retirement. Graduated projects become Top-Level Projects (TLPs) with Project Management Committees (PMCs). Sustainability within ASF is tied to cultivating and maintaining self-governing, collaborative communities.

\textbf{Eclipse Foundation Incubators (EFI).}
\label{EFI}
The EFI \cite{eclipse} operates under the Eclipse Development Process (EDP) \cite{17_1eclipsefoundation}, guiding projects from proposal to retirement. After formal proposal review and acceptance, projects enter incubation (typically one to two years \cite{24_eclipse_incubation}), focusing on establishing legally vetted codebases, cultivating diverse contributors, releasing Eclipse-standard software, and demonstrating independent governance capabilities.

\begin{figure}[]
\includegraphics[width = \linewidth, height=5cm]{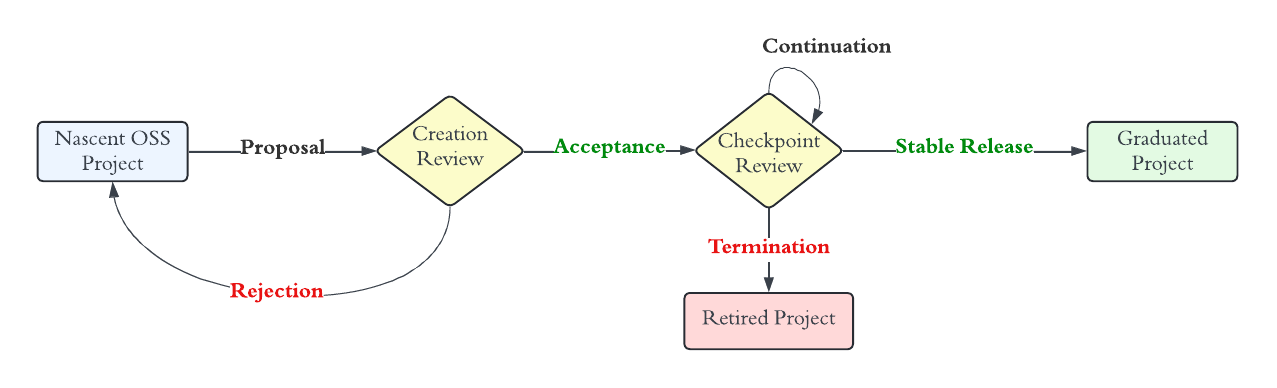}

\caption{Incubation Process of Eclipse Foundation}
\label{incubation_eclipse}
\end{figure}

Figure~\ref{incubation_eclipse} illustrates the lifecycle stages: \textit{Incubation}, \textit{Mature}, and \textit{Top-Level}. Advancement requires meeting EDP criteria \cite{17_3eclipsefoundation}; exceptional projects may bypass Mature status for direct Top-Level promotion. Projects failing to meet criteria are moved to Archived status, analogous to Apache retirement.

Unlike ASF's mailing list mandate, Eclipse requires open communication channels, Eclipse IP policy adherence, and EDP best practices. Project Management Committees (PMCs) provide technical direction and oversight \cite{17_2eclipsefoundation}. Sustainability hinges on fostering independent, transparent development culture while delivering reliable, standards-compliant software.

\begin{figure}[]

\includegraphics[width = \linewidth, height=7cm]{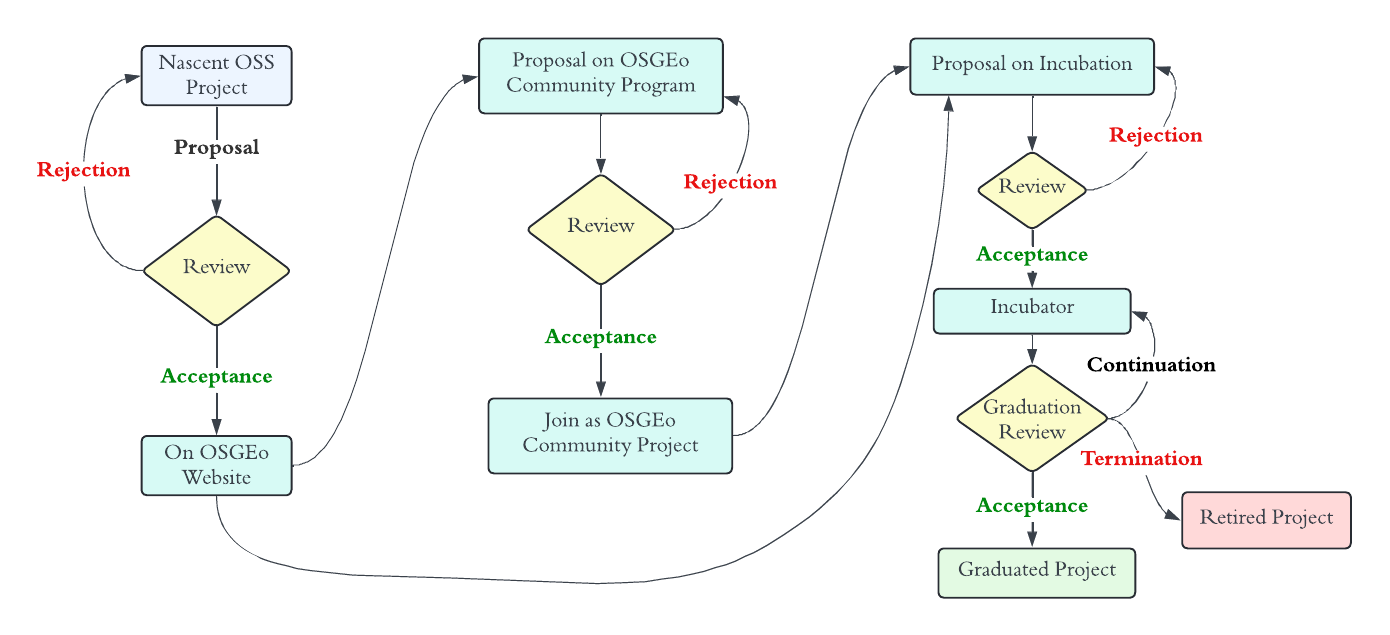}

\caption{Incubation Process of OSGEo Foundation}
\label{incubation_osgeo}

\end{figure}

\textbf{OSGeo Foundation Incubators (OFI).}
\label{OSGEo}
The OSGeo Foundation (OF) \cite{osgeo} supports geospatial technology OSS projects, operating at the intersection of open-source development and Geographic Information Systems (GIS) research. OF employs a multi-tiered incubation pipeline (Figure~\ref{incubation_osgeo}) with flexible entry points.

Projects may enter through the Community Projects Program, receiving lightweight guidance on governance, licensing, and community development. Projects demonstrating adoption potential proceed to formal incubation under the OSGeo Incubation Committee, receiving structured mentorship, reviews, and compliance assessment. Successful projects undergo Graduation Review for official OSGeo project status with long-term support. This tiered structure offers greater adaptability than monolithic models like ASF, accommodating varied project maturity levels.

\textbf{GitHub.}
GitHub \cite{GitHub} is a platform for hosting and collaborating on OSS projects, lacking centralized governance or formal incubation \cite{18_GitHub_distribution}. Any individual or organization can create projects \cite{19_GitHub_code_of_conduct}, yielding diverse ownership models. The fork-and-pull-request workflow \cite{25_rahman2014insight} emphasizes contribution ease while giving owners significant control, unlike foundation consensus-driven models. 

GitHub provides optional governance tools (organization accounts, teams, permissions) but no standardized sustainability definition. This flexibility has fueled rapid OSS growth but also high failure rates due to absent mentorship, legal backing, and governance frameworks available in foundations \cite{27_coelho2018identifying}.

\section{Research Questions}

Sociotechnical networks have proven effective in modeling emergent properties of software development ecosystems, such as collaboration patterns, project health, and community structure. Building on this foundation, we hypothesize that sustainability across OSS projects can also be modeled through sociotechnical networks.
This motivates our central inquiry: 


\newcommand{\RqOne}{RQ1:
Does a unified model of sustainability based on sociotechnical network features exist between the ASF, EF, and OF foundations? Can sustainability be predicted across foundations—i.e., can a model trained on one foundation predict sustainability in another?}

\noindent
\fbox{%
  \parbox{\linewidth}{%
    \RqOne
  }%
}

Analyzing cross-foundation models not only reveals performance differences but also helps identify which sociotechnical features are foundation-specific and which are transferable. This insight is crucial for practitioners seeking to make informed, data-driven decisions in project planning and implementation.


While foundation-specific sustainability models effectively capture the collaboration and activity dynamics unique to each ecosystem, they cannot be applied indiscriminately across projects originating outside those ecosystems. To extend sustainability prediction to new or unaffiliated projects, an essential step is to first determine which software foundation a project most closely resembles in its sociotechnical structure. This “routing” ensures that each project is evaluated by the sustainability model whose governance, collaboration, and activity patterns best align with its own.

\newcommand{\RqTwo}{RQ2: Based on sociotechnical network features, can we identify the software foundation to which an open-source project is most analogous, and, once routed, can the corresponding foundation-specific sustainability model effectively forecast its sustainability outcome?}

\noindent
\fbox{%
  \parbox{\linewidth}{%
    \RqTwo
  }%
}

The concept of sustainability has been popularized by software foundations. But in the absence of requirements for performance, it is not very useful for projects outside of them, e.g., for projects that only exist on GitHub. For these, prior work have used quantitative metrics, based on digital traces, and have mostly called them \emph{project success}. Here we ask if models of (e.g., success) and (e.g., sustainability) are interchangeable.

\newcommand{\RqThree}{RQ3: What is the relationship between models of OSS project success and OSS project sustainability? Can our sustainability models be used with projects outside of foundations?}

\noindent
\fbox{%
  \parbox{\linewidth}{%
    \RqThree
  }%
}




To further interpret these models, we investigate which sociotechnical features are most influential in determining sustainability outcomes. Different foundations may prioritize or benefit from different community or development dynamics.

\newcommand{\RqFour}{RQ4: Which sociotechnical features are most predictive of sustainability in the best-fit models for different foundations and ecosystems?}

\noindent
\fbox{%
  \parbox{\linewidth}{%
    \RqFour
  }%
}




\begin{figure*} [!ht]
\centering
\includegraphics[width = 0.95\linewidth, height=14cm]{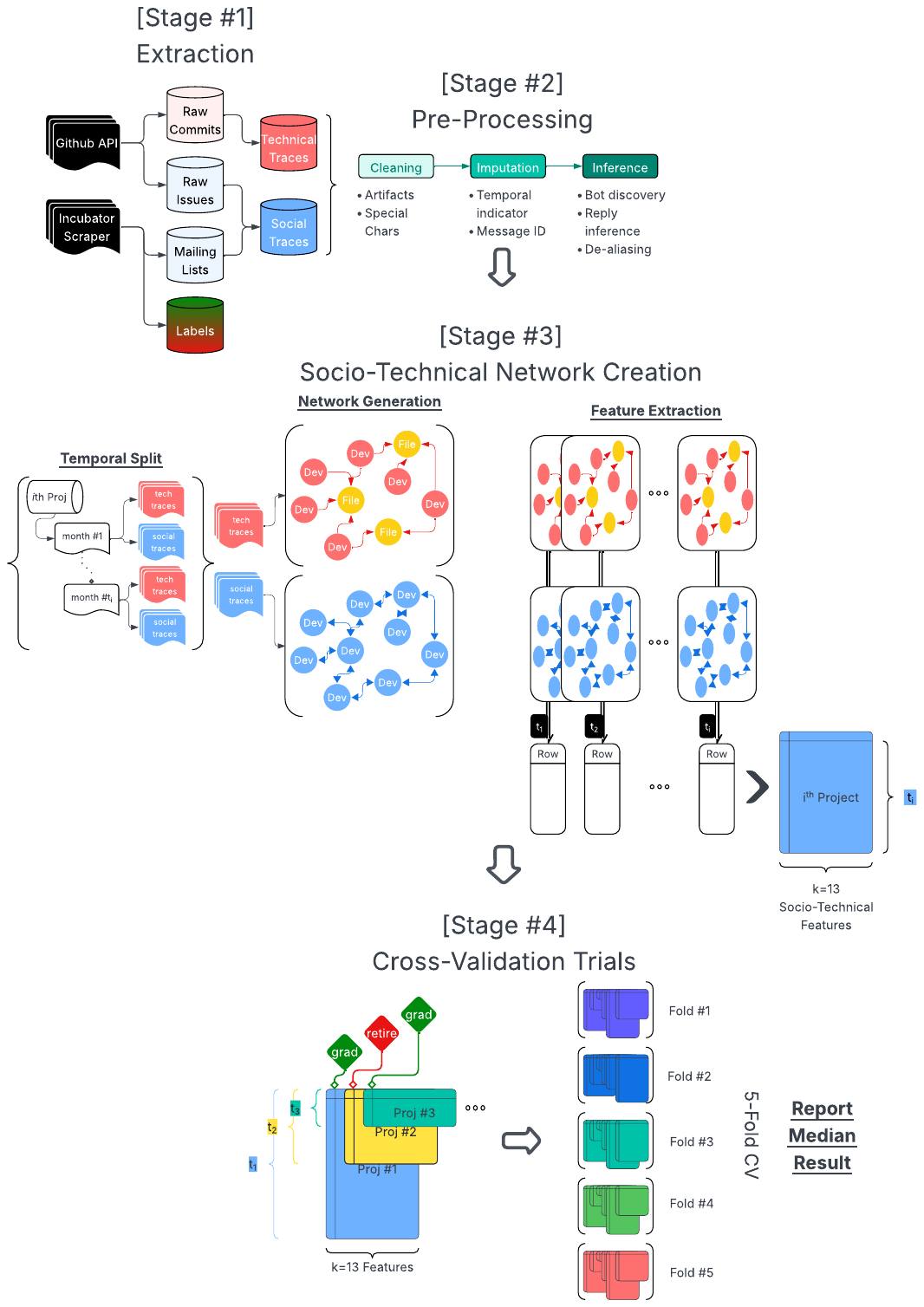}
\caption{Methodology of the proposed study}
\label{study-methodology}
\end{figure*}


\section{Methodology}

\subsection{Dataset Gathering}

Table~\ref{tab:dataset} summarizes the composition of our final dataset across all four ecosystems. For each ecosystem, the table reports the total number of projects identified, the number excluded due to active incubation status (i.e., lacking a definitive sustainability outcome), the number excluded due to absence of technical traces, the number excluded due to missing labels, and the final count of graduated and retired (or successful/unsuccessful) projects used for modeling. The subsections below describe the data collection process for each ecosystem in detail.

\subsubsection{ASF Data}
For ASF, we utilized the comprehensive dataset published by Yin et al.~\cite{3_yin2021apache}, which captures developer coding and communication activities across 329 projects in the Apache Software Foundation Incubator (ASFI). The ASFI process results in one of two terminal outcomes for each project: graduation or retirement. Of the projects in the dataset, 223 had graduated, 69 were retired, and 37 were still in incubation at the time of data collection. As shown in Table~\ref{tab:dataset}, the 37 incubating projects were excluded as they lack definitive sustainability outcomes, and a further 30 projects were excluded due to the absence of technical traces—leaving a final set of 262 
projects (205 graduated, 57 retired). In their paper, Yin et al. focused on each project's incubation period—i.e., from the time a project entered the ASFI until its graduation or retirement—rather than its full development lifespan. We also leveraged the same time spans for each project for our study.

\subsubsection{EF Data}
To collect data from the EF, we developed a Rust-based mining tool~\cite{OSSPREY_OSS_Scraper_Tool} tailored to the Eclipse Project Management Infrastructure (PMI). The tool programmatically accesses project metadata, clones Git repositories from EF's GitLab instance, and scrapes mailing list archives and issue trackers. We identified a total of 511 EF incubator projects: 194 graduated, 183 retired, and 134 still in incubation. Since incubating projects lack definitive sustainability outcomes, the 134 incubating projects were excluded from model training and evaluation. Additionally, 216 projects were excluded due to the absence of technical traces, as these projects hosted their code on platforms other than GitHub and could not be mined by our tool. As summarized in Table~\ref{tab:dataset}, this yielded a final set of 161 projects (142 graduated, 19 retired). To maintain consistency with our treatment of the ASF dataset, we restricted our analysis to the incubation phase of each project, starting from the date of proposal acceptance to the point of graduation or retirement.

\subsubsection{OF Data}
We used the same Rust mining tool~\cite{OSSPREY_OSS_Scraper_Tool} to collect data from the OF. The tool was adapted to interface with OSGeo's project metadata sources, including its website and community wiki, and to gather technical traces from GitHub repositories. We identified 68 projects in total: 25 had successfully graduated through the incubation process, 4 were actively in incubation, and 19 were listed as community projects. The remaining 20 projects were either withdrawn or lacked sufficient digital traceability for analysis. As shown in Table~\ref{tab:dataset}, after excluding the 4 incubating projects and the 25 projects lacking technical traces or labels, the final OSGeo dataset comprises 20 projects (13 graduated, 7 retired). As with ASF and EF, we focused exclusively on each project's incubation phase, reflecting the period in which the foundation typically makes 
sustainability-related evaluations.

\subsubsection{GitHub Data}
Unlike ASF, EF, or OF, GitHub lacks a centralized governance or incubation mechanism, making it more difficult to assign sustainability labels. While prior studies have attempted to define sustainability using intrinsic metrics like duration of activity~\cite{4_xiao2023early}, we adopted the curated dataset from Joblin et al.~\cite{6_joblin2022successful}, which classifies GitHub projects as either successful or unsuccessful based on long-term outcomes. As shown in Table~\ref{tab:dataset}, of the 32 projects in this dataset, 11 were excluded due to the absence of technical traces, yielding a final set of 21 projects (5 successful, 16 unsuccessful). To match the temporal framing used in the foundation 
datasets, we followed Joblin et al.'s approach of selecting the most active initial portion of a project's lifecycle. We considered project activity only during the early development phase, prior to long-term scaling or decline, thereby reducing confounding factors introduced by differing 
project ages.

\subsubsection{Project Length Distribution Across the Ecosystems}
\label{length_distribution}
\begin{figure}[]
\includegraphics[width = \linewidth, height=5cm]{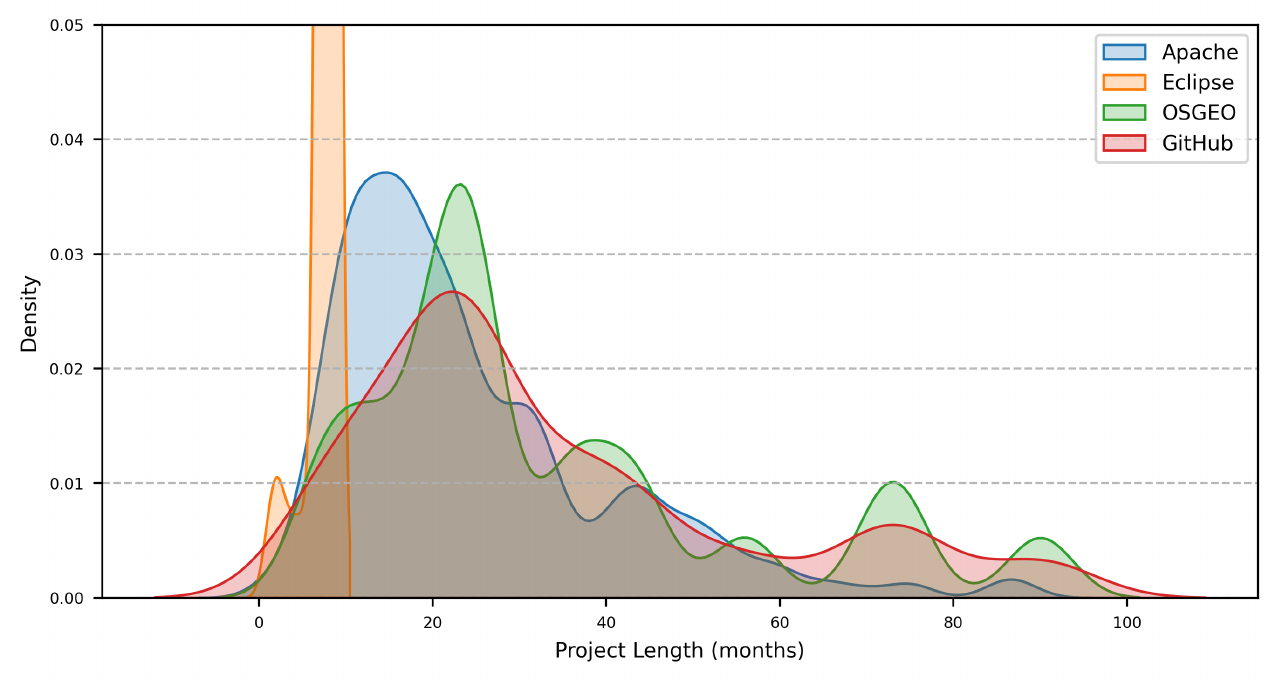}
\caption{Distribution of Project Length Across Different Foundation/Ecosystems}
\label{fig:project_length_distribution}

\end{figure}

Projects in the different ecosystems vary substantially in duration. Figure~\ref{fig:project_length_distribution} shows the distribution of project lengths (in months) across four ecosystems: AF, EF, OF, and GitHub. The x-axis represents project duration, while the y-axis reflects density, truncated at 0.05 to allow all foundations to be visualized on a common scale. Notably, incubation periods vary significantly across foundations. EF projects exhibit a sharply peaked distribution, reflecting a standardized and shorter incubation process. In contrast, AF and OF show broader and often multimodal distributions, indicating diverse incubation trajectories. GitHub, lacking a formal incubation framework, displays a wider and flatter distribution, capturing the organic and varied nature of project lifecycles. This diversity highlights the importance of tailoring sustainability modeling to the structural characteristics of each ecosystem.

A consistent principle across all datasets was to restrict modeling to the early and most formative stages of each project's lifecycle. For foundational projects (ASF, EF, OF), we analyzed only the incubation period—the window during which projects are actively evaluated for sustainability outcomes such as graduation or retirement. For GitHub, we targeted the early peak of contributor activity. This decision was grounded in our goal: \textbf{to determine whether it is possible to predict a project's sustainability outcome based solely on its observable behavior during its evaluation phase.} 

Modeling data from the full project lifespan could introduce label leakage—for instance, discontinued projects would show long periods of inactivity, leading models to trivially predict retirement. By focusing on periods when developers are actively working toward sustainability goals and when project outcomes are uncertain, we ensure that our models are trained on realistic, decision-relevant data.

\begin{table*}[t]
\centering
\caption{Summary of Project Attrition and Final Dataset Composition Across Apache, Eclipse, OSGeo, and GitHub Ecosystems, After Filtering for Incubation Status, Technical Trace Availability, and Label Completeness}
\label{tab:dataset}
\begin{tabular}{
p{1.6cm}  
p{1.5cm}  
p{1.7cm}  
p{1.6cm}  
p{1.3cm}    
p{1.4cm}  
p{1.4cm}  
}
\toprule
Incubator &
Identified &
Incubating &
No Tech Traces &
No Labels &
\multicolumn{2}{c}{Final} \\
\cmidrule(lr){6-7}
 & & & & & Grad & Retired \\
\midrule
Apache & 329 & 37 & 30 & 0 & 205 & 57 \\
Eclipse & 511 & 134 & 216 & 0 & 142 & 19 \\
OSGeo & 68 & 4 & 25 & 19 & 13 & 7 \\
GitHub & 32 & 0 & 11 & 0 & 5 & 16 \\
\bottomrule
\end{tabular}
\end{table*}


\subsection{Data Pre-Processing}
\textbf{Source File Cleaning}. To ensure only technical contributions were tracked, we excluded non-source code files from our study. We leveraged a comprehensive file available on GitHub \cite{gist} containing file extensions frequently associated with source files. Utilizing the defined list of source-code extensions, we removed non-source-code files, capturing a more precise representation of the project's technical development activities and collaborations.

\textbf{Bot Participation Removal.} 
Bots are frequently used in OSS projects to automate routine or repetitive tasks \cite{7_wessel2018power}. While they are valuable for project maintenance, their inclusion in sociotechnical analyses can distort our understanding of actual developer behavior. Bot-generated activity can introduce noise into social and technical networks, potentially diluting signals of genuine human collaboration \cite{8_wessel2021don}.

\begin{figure}[!ht]
\centering
\includegraphics[width = 0.95\linewidth, height=14cm]{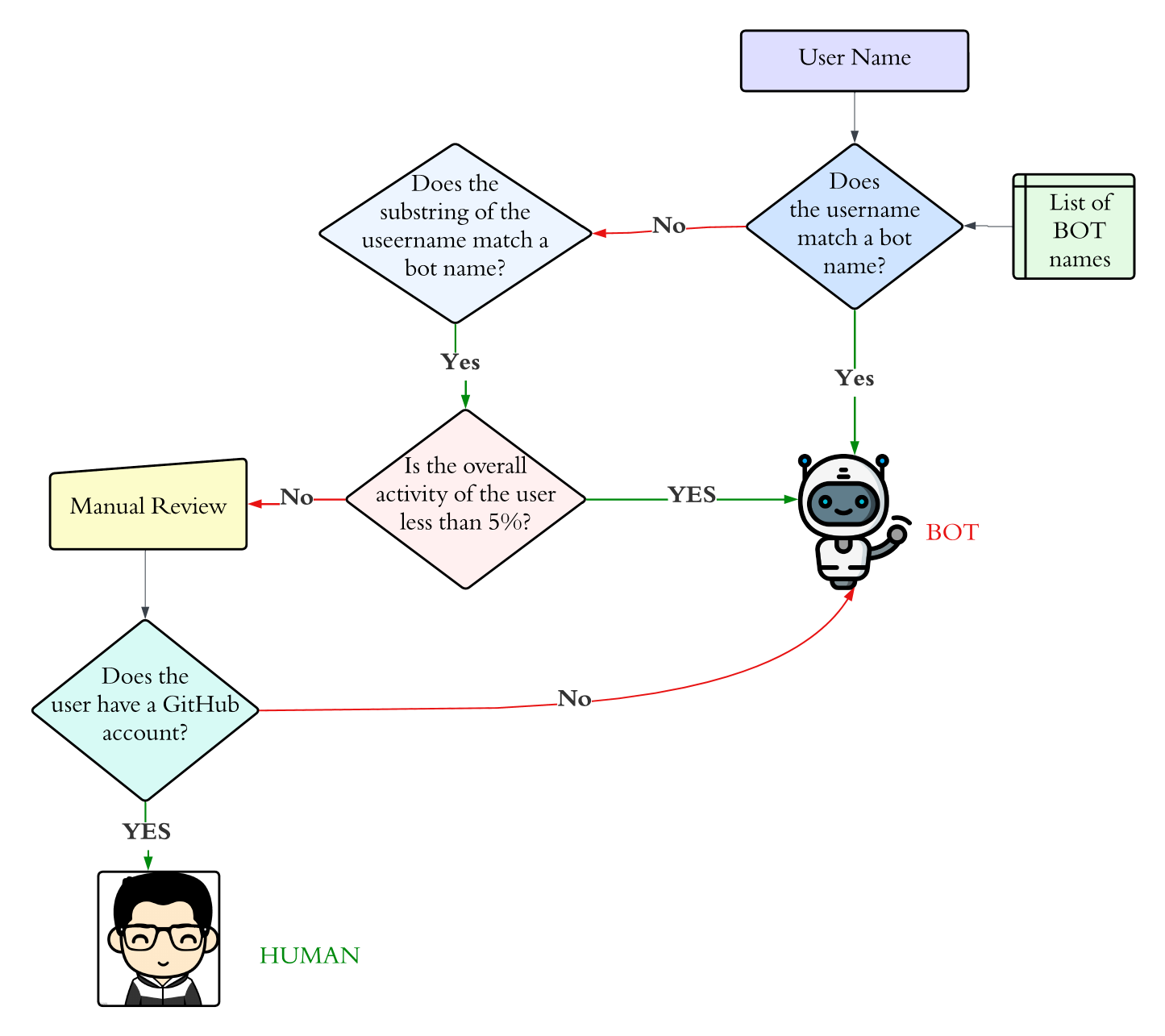}
\caption{Bot removal process}
\label{bot-removal}
\end{figure}

The bot removal process is illustrated in Figure \ref{bot-removal}. Users are flagged as bots based on two main criteria. First, a substring match checks whether the username contains common bot-related terms (e.g., "bot", "butler") or project-specific identifiers (e.g., "tensorflow-bot"). Second, the process applies two follow-up conditions: (i) whether the username shares substrings with known bots, and (ii) whether the user's overall activity accounts for less than 5\% of project events. If both conditions are met, the account is automatically flagged as a bot. If either fails, the account is marked for manual review. During manual review, we check for the existence of a GitHub profile. If a valid profile exists, the user is labeled as 'Human'; otherwise, the user is labeled as a 'Bot'. Once identified, all social and technical activities associated with bot accounts were excluded from the dataset.

\begin{figure*}[]
\centering
\includegraphics[width =12cm, height=6cm]{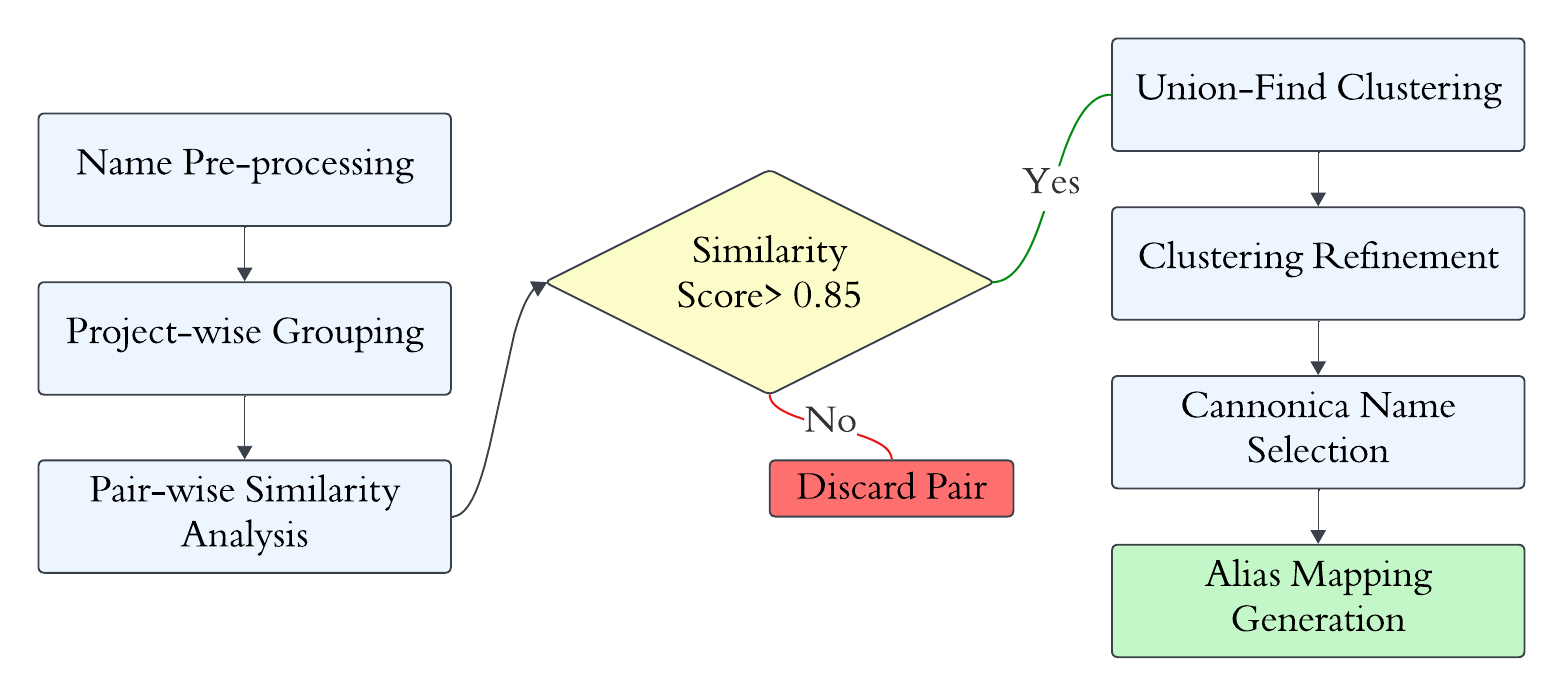}
\caption{Alias resolution pipeline used for de-aliasing OSS contributor identities}
\label{de-aliasing_email}
\end{figure*}

\textbf{De-aliasing User Accounts}. It is common to find the same individual under multiple account names. This aliasing may occur because developers use different accounts to separate personal and professional identities, participate anonymously, or follow organization-specific naming policies \cite{9_vasilescu2015gender}. Aliasing complicates the accurate tracking of individual contributions, inflates the number of contributors, and introduces noise in sociotechnical analyses. To mitigate, we implement a comprehensive \textit{de-aliasing pipeline}, as illustrated in Figure~\ref{de-aliasing_email}. This multi-stage process systematically detects and merges different representations of the same contributor, resulting in cleaner, more accurate identity mappings.
 
The pipeline begins with a \textit{Name Pre-processing} module, where raw names are cleaned and normalized. This step removes extraneous information such as email domains (e.g., \texttt{@example.com}), text within brackets (often roles or timestamps), and special symbols like dollar signs or punctuation marks. For instance, names like ``john.doe@company.com'' and ``John Doe (Developer)'' are reduced to comparable formats—``john doe''—to improve matching consistency.

Next, in the \textit{Project-wise Grouping} step, the algorithm processes each project independently. This is important because similar names across unrelated projects (e.g., ``Alex Kim'' in Project A and Project B) may belong to different individuals. By restricting analysis within project boundaries, we reduce the likelihood of false matches.

The third module, \textit{Pair-wise Similarity Analysis}, computes the similarity between every pair of developer names within a project. We apply the Jaro-Winkler \cite{110_rozinek2024fast} string similarity metric, which is effective for detecting variations due to typos, abbreviations, or informal naming (e.g., ``Mike'' vs. ``Michael''). To avoid false positives from shared first or last names, we include a validation step that ensures sufficient distinction across name segments.

The fourth stage is a \textit{Threshold-Based Decision Module}, where we determine whether two names are similar enough to be considered aliases. We use a high similarity threshold of 0.85, meaning that only names with a high degree of similarity are grouped. Pairs below this threshold are discarded to prevent incorrect merges. Pairs that pass this test enter the \textit{Union-Find Clustering} module. This phase groups names not just through direct similarity but also through transitive relationships. For example, if ``J. Smith'' matches ``John Smith'', and ``J. Smith'' also matches ``jsmith'', then all three are grouped together, even if ``John Smith'' and ``jsmith'' were not directly matched. This allows us to identify alias chains that reflect how people naturally vary their usernames.

Once initial clusters are formed, a \textit{Multi-Stage Refinement} step filters out weak or noisy links. Here, we remove names from clusters if their average similarity to others is low, apply segment validation checks again, and eliminate suspiciously large or tiny clusters—since over-clustering or singleton clusters often indicate errors.


Each refined cluster is then assigned a \textit{Canonical Name}—the most complete and representative version of a contributor’s identity. We prefer two-part names (e.g., ``Jane Doe'') that are properly capitalized and longest in character count, assuming they are less likely to be abbreviations or artifacts. Finally, the \textit{Alias Mapping Generation} module links every alias in a cluster to the canonical name and finds the most representative email address (usually from commit metadata). The result is a project-specific JSON mapping that can be easily integrated into downstream analyses.

The above-mentioned de-aliasing procedure allows us to unify fragmented contributor records, generate more realistic developer networks, and ensure that metrics related to participation, collaboration, and sustainability reflect actual individuals, not duplicated aliases.

\textbf{Reply Inference.} 
In datasets from platforms like EF, OF, and GitHub, explicit reply links—indicating exactly who responded to whom within an issue or email thread; are often absent due to the unstructured nature of these conversations. However, such reply information is critical for constructing accurate social networks that reflect communication pathways among developers. 

\begin{figure}[]

\includegraphics[width = \linewidth, height=7cm]{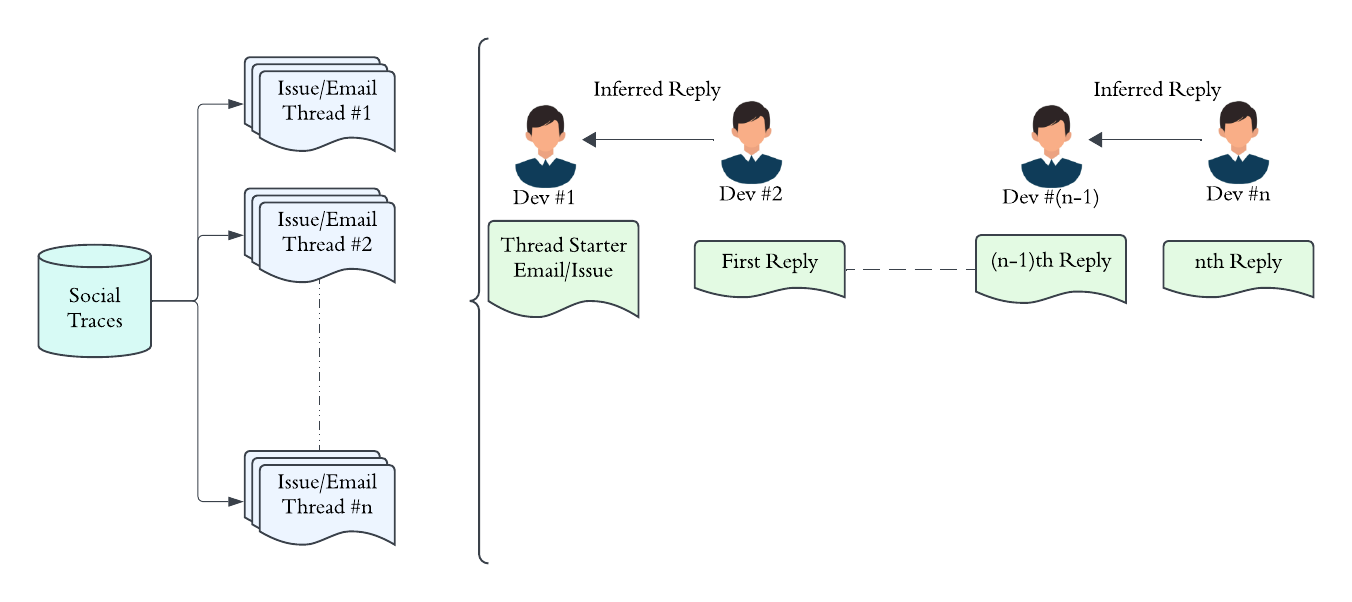}

\caption{Reply Inference Process}
\label{reply_inference}
\end{figure}

To infer reply relationships, we relied on structural and temporal cues present in comment threads. The reply inference process is illustrated in Figure \ref{reply_inference}. For each issue or email thread, communication is inferred based on the sequence and timing of user replies. When a second user (user~\#2) posts a response following the original thread creator (user~\#1), we treat this as a social interaction between the two users. For the $i^\text{th}$ reply in the thread, we assume the respondent is replying to the $(i-1)^\text{th}$ user, unless the reply is to themselves. These inferred interactions are used to construct edges in the project's social network graph.

\subsection{Social \& Technical Networks}
We generated social and technical networks for each project in the dataset. Social networks capture interactions between contributors, as nodes, where a directed edge exists between A and B if B has responded to a communication by A.
A technical network captures co-contribution to the same file between developers A and B in a given period, as an undirected edge between nodes A and B.

For EF and OF projects, we cloned all git repositories and obtained technical trace data for each repository via traversal with the miner tool. We then aggregated traces over all the project's repositories to get project-level interactions. Social traces were similarly collected via the miner tool for issues and email communications. Leveraging this trace data, sociotechnical networks were generated on a monthly basis as previously described, thus giving a temporally varying representation of developers' social and technical collaboration for each project.

Similarly, we leverage pre-compiled data from Yin et al.~\cite{3_yin2021apache} for ASFI projects and Joblin et al.~\cite{6_joblin2022successful} for GitHub projects to generate sociotechnical networks.

After generating the networks, we identified that some projects lacked technical activities. This absence is because these projects host their technical traces on platforms other than GitHub. Since software can not exist without any technical activity, we excluded these projects from our final dataset. The final dataset consists of 262 projects from the ASF (Graduated: 205, Retired: 57), 161 projects from the EF (Graduated: 142, Retired: 19), 20 projects from OF (Graduated: 13, Retired: 7) and 21 projects from GitHub (Successful: 5, Failure: 16). The complete list of all projects with their graduation outcomes is available at:
\href{https://docs.google.com/spreadsheets/d/e/2PACX-1vTcdskobmxZWG40ExEAu7QrtnDfYgS08OFX_MkTfEMUxgNLnc82WgvDHixaxLvqIojBHVm0ypU5X9kl/pubhtml}{this project list}.

\raggedbottom

\subsection{Features}
\textbf{Feature Selection}. To develop the forecasting models, we primarily used the features from the recent work of Yin et al. \cite{1_yin2023self}. Their research examined episodic changes in sociotechnical aspects of ASF Incubator Projects, considering 16 features: ten sociotechnical, three institutional agents, and three institutional resistance features. Given our focus on predicting sustainability through temporal traces of sociotechnical features, we incorporated all ten sociotechnical features from their study. Additionally, we included three other features based on thorough discussions and consideration of the underlying theories of sociotechnical systems measures. All selected features were additionally  checked for collinearity.

The social features from the prior paper include 1) the number of unique active developers \emph{s\_num\_nodes} in social networks; 2) average clustering coefficient \emph{s\_avg\_clustering\_coef} describing the linkage of a node to its neighbors; 3) graph density \emph{s\_graph\_density} measuring network connectivity; 4) total number of disconnected components \emph{s\_num\_component} in the social network; 5) weighted mean degree \emph{s\_weighted\_mean\_degree} of the social networks. Moreover, we included another social feature 6) \emph{s\_net\_overlap}, denoting the number of developers consistently active in social networks (active in the current as well as the previous month). 

On the technical side, the included features from the paper are 7) the density of the technical network \emph{t\_graph\_density}; 8) the degree of collaborative behaviors measured as developers per file node (t\_num\_dev\_per\_file); 9) number of unique developers in technical networks (t\_num\_dev\_nodes); 10) the number of unique coding files (t\_num\_file\_nodes); 11) degree of multitasking represented by files per developer node (t\_num\_file\_per\_dev). Like social networks, here also we included a feature \emph{t\_net\_overlap}, which counts the number of consistent developers in the technical network.

Additionally, we incorporated a mixed feature 13) \emph{st\_num\_dev} that counts developers contributing to both technical and social aspects of the project, bridging the gap between these two dimensions. The final list of 13 selected features allows us to capture the key aspects of both the technical and social activities of a project. The descriptive statistics are given in Table \ref{DataStats}.


\begin{table*}[ht]
\centering
\caption{Statistics of socio-technical features across ASF, EF, OF, and GitHub ecosystems.}
\label{DataStats}
\resizebox{\textwidth}{!}{\renewcommand{\arraystretch}{0.85}
\begin{tabular}{@{}lcccccccc@{}}
\toprule
\textbf{Feature Name} & 
\multicolumn{2}{c}{\textbf{Apache}} & 
\multicolumn{2}{c}{\textbf{Eclipse}} & 
\multicolumn{2}{c}{\textbf{OSGeo}} & 
\multicolumn{2}{c}{\textbf{GitHub}} \\
\cmidrule(lr){2-3} \cmidrule(lr){4-5} \cmidrule(lr){6-7} \cmidrule(lr){8-9}
 & Mean & Std. Dev. & Mean & Std. Dev. & Mean & Std. Dev. & Mean & Std. Dev. \\
\midrule
\texttt{s\_avg\_clustering\_coef} & 0.1864 & 0.1241 & 0.0800 & 0.1300 & 0.0027 & 0.0137 & 0.0112 & 0.0370 \\
\texttt{s\_graph\_density}        & 0.2179 & 0.1753 & 0.3800 & 0.3900 & 0.0086 & 0.0339 & 0.1261 & 0.2356 \\
\texttt{s\_net\_overlap}          & 0.1503 & 0.0716 & 0.0900 & 0.0800 & 0.0000 & 0.0000 & 0.0343 & 0.0444 \\
\texttt{s\_num\_component}        & 11.7669 & 5.6310 & 2.4100 & 1.4800 & 0.1524 & 0.7036 & 40.0814 & 29.6974 \\
\texttt{s\_num\_nodes}            & 38.5252 & 15.8392 & 6.1500 & 3.5700 & 0.5668 & 2.6579 & 73.9179 & 174.6009 \\
\texttt{s\_weighted\_mean\_degree} & 11.1416 & 6.0114 & 3.9700 & 3.7900 & 0.1884 & 1.2056 & 0.7830 & 2.2733 \\ \hline
\texttt{st\_num\_dev}             & 2.6384 & 1.4692 & 0.8500 & 0.7100 & 0.0070 & 0.0442 & 0.2307 & 0.3726 \\ \hline
\texttt{t\_graph\_density}        & 0.3373 & 0.2433 & 0.3600 & 0.3100 & 0.4019 & 0.2346 & 0.3223 & 0.3343 \\
\texttt{t\_net\_overlap}          & 0.0647 & 0.0646 & 0.0400 & 0.0400 & 0.0741 & 0.0703 & 0.0689 & 0.0793 \\
\texttt{t\_num\_dev\_nodes}       & 5.2472 & 2.6180 & 3.0800 & 1.5400 & 4.4770 & 1.9215 & 4.9551 & 3.8464 \\
\texttt{t\_num\_dev\_per\_file}   & 1.0217 & 0.4266 & 0.7700 & 0.3300 & 1.1397 & 0.3730 & 0.8517 & 0.6791 \\
\texttt{t\_num\_file\_nodes}      & 271.2600 & 269.3800 & 194.1800 & 288.6200 & 219.5822 & 410.7125 & 200.3800 & 118.2700 \\
\texttt{t\_num\_file\_per\_dev}   & 59.1244 & 67.0070 & 65.8046 & 129.4125 & 57.3980 & 121.8164 & 47.8547 & 55.9677 \\
\bottomrule
\end{tabular}
}
\end{table*}

\textbf{Feature Normalization}. When dealing with datasets comprising projects of varying scales and characteristics, a crucial step before training the model is data normalization \cite{14_hutter2019automated}. This process mitigates the risk of exploding gradients problem which can severely impact model performance and stability \cite{109_philipp2017exploding}. Due to the unique characteristics of the dataset, we found the traditional normalization techniques to be inadequate. For example: Min-Max Normalization \cite{21_mazziotta2022normalization}, which scales features to a fixed range (typically 0 to 1) was not suitable for our study as it can introduce a vanishing gradient problem in features with large variances across different projects. This limitation is particularly pronounced where activity levels can vary dramatically between small and large projects. The Z-Score Normalization \cite{22_fei2021z}, which standardizes features based on their mean and standard deviation, is also not applicable, as it assumes the data is normally distributed, which does not hold for OSS projects, as seen above, and in prior papers.

Instead, we employed a project-scale-aware normalization. For project $p$ at month $t$, let $D_p^{(t)}$ denote the union of developers active in social or technical networks. For all features \textit{except} developer counts (\texttt{s\_num\_nodes}, \texttt{t\_num\_dev\_nodes}), we normalize: $f_{\text{norm}}^{(t)} = f^{(t)} / |D_p^{(t)}|$. Developer count features are excluded from normalization to preserve their discriminative capacity; normalizing them would create degenerate features constrained to $[0, 1]$. This approach controls for project scale while retaining the raw signal of team size variation across projects.

\subsection{Model Development}

In this study, we propose three neural network architectures, \textbf{Bi-LSTM}, \textbf{Transformer}, and \textbf{Bi-DLSTM} to forecast the sustainability of OSS projects, as a single model architecture might not be capable of capturing the hidden patterns and characteristics of multiple incubators. As the data of our study is longitudinal time series data, we proposed bidirectional models, which can process the input sequence in both forward and backward directions. This is required to capture dependencies and relationships in the data that may occur both before and after a given time point, which can improve the model's ability to understand and predict complex patterns over time. 

The description of each architecture is provided as follows:
\textbf{Bi-LSTM}. Bidirectional LSTM (Bi-LSTM) captures long-range dependencies in both directions, which enables it to understand the context of the entire sequence \cite{12_BiLSTM}. The proposed model architecture consists of seven interconnected layers. The architecture begins with a bidirectional LSTM layer, the output of which is subsequently fed into an attention mechanism. The attention mechanism selectively weighs the importance of different time steps, enabling the model to concentrate on critical segments of the data that significantly impact the target variable \cite{44_niu2021review}. The first fully connected layer applies a linear transformation to the input, adjusting its dimensionality. This is immediately followed by layer normalization, which standardizes the activations of the previous layer to stabilize and accelerate the training process. The Rectified Linear Unit (ReLU) \cite{agarap2018deep} activation function is then applied to introduce non-linearity into the model, enabling it to learn more complex and abstract features. To mitigate overfitting and enhance the model's generalization, a dropout layer is added next, which randomly deactivates a subset of neurons during training. After the first sequence of fully connected, normalization, activation, and dropout layers, we implemented a residual connection \cite{38_residual} in the architecture. The residual connection takes the output from the first set of layers and adds it to the output of the next layer. This design helps in two ways: it facilitates the network's learning by allowing gradients to flow smoothly during backpropagation, and second, it helps prevent the problem where the network struggles to learn from data that has passed through many layers. Next, the processed output from the first sequence of layers, incorporated with the residual connection, is then fed into another sequence of layers containing fully connected, normalization, activation, and dropout layers. This sequence of layers is added to refine the extracted features further. After that, we consolidated the learned representations of the model and produced the final variable by utilizing a fully connected layer. As an activation function of the final layer, Softmax has been used, which generates a probability distribution across the classes. 

\textbf{Transformer}. The Transformer architecture replaces recurrent units with self-attention mechanisms, allowing it to model long-range dependencies without sequential processing \cite{vaswani2017attention}. It uses multi-head attention and position-wise feed-forward layers to capture contextual relationships across the entire input sequence in parallel. In our setup, the Transformer encoder layer replaces the first recurrent layer, while the rest of the architecture remains consistent with the Bi-LSTM baseline. This design enables efficient parallelization, often leading to improved performance in sequence modeling tasks.

\textbf{Bi-DLSTM}. Bi-Dilated-LSTM incorporates dilated connections in addition to bidirectional processing, allowing it to capture a broader range of temporal dependencies \cite{10_1dilated, 10_2dilated}. This architecture can effectively increase the receptive field without increasing the number of parameters, making it particularly useful for tasks requiring multi-scale temporal information. Instead of single-headed attention, we used multi-headed attention for this architecture. Other than that, we kept the rest of the architectural design identical to that of the Bi-LSTM and Transformer models.

\textbf{Architectural Design Considerations}. The architectures of the models were proposed through an extensive series of experiments, employing a methodical trial-and-error approach \cite{45_goodfellow2016deep, 46_sosna2010business}. Throughout this process, we meticulously observed the training loss across each epoch to make decisions in structuring the network, for example: adding dense, normalization, and dropout layers. As observing the evidence of gradual learning, we increased the network's complexity by adding more neurons and layers. Conversely, given the risk of overfitting, particularly in the absence of validation set, we incorporated dropout layers as a preventive regularization measure. Although training loss continued to decline, this alone was insufficient to guarantee generalization. Dropout mitigates overfitting by introducing stochasticity during training, forcing the model to learn more robust, distributed representations rather than memorizing patterns in the training data \cite{36_hawkins2004problem}. To add flexibility in utilizing or bypassing certain layers of the architecture, depending on the complexity of the features being learned, we included residual connections. These connections also allow the model to learn residual functions by creating shortcuts for gradient flow during backpropagation. In parallel with the architectural design process, we employed a grid search algorithm \cite{37_grid} to systematically explore the hyperparameter space, involving the parameters: learning rate, dropout rate, batch size, number of attention heads, and the hidden size of the LSTM layers. By following this process, we finally selected the combination value of the hyperparameters for each trial combination.

\subsection{Model Training}

Next, we trained and evaluated our proposed models on various training/testing data combinations, drawn from different foundations. These fall into three broad categories: (1) \textit{intra-foundational trials}, (2) \textit{cross-foundation trials}, and (3) \textit{GitHub trials}.
A notable challenge in our training setup was the presence of significant class imbalance, with the majority of projects labeled as “graduated.” To mitigate this issue, we adopted \textit{focal loss} \cite{111_mukhoti2020calibrating}, as the loss function, which down-weights easy examples and focuses training on hard, misclassified samples. This choice was particularly effective for our binary classification setting and helped reduce the bias introduced by skewed class distributions. 
Each sample in our dataset represents an individual project, formatted as a time-indexed tensor or data frame. Regardless of project length, the model predicts a single binary outcome, graduated or retired, based on the entire incubation window. All proposed architectures were designed to handle variable-length sequences as input.

Training was carried out using an adaptive learning rate scheduler, which dynamically adjusted the learning rate based on the training loss to accelerate convergence. Each model was trained for up to 200 epochs, with early stopping \cite{48_prechelt2002early} activated if no improvement was observed in training loss for 10 consecutive epochs. The best-performing model weights were saved during training for final evaluation. To address potential exploding gradient issues, we incorporated gradient clipping \cite{gczhang2019gradient} after each backward pass. These measures collectively ensured that the models converged efficiently while avoiding overfitting and instability during training.

\subsection{Model Interpretability}
We utilized the SHapley Additive exPlanations (SHAP) technique \cite{34_van2022tractability} to determine the importance of each of the features influencing the desired outcome of 'graduation' across the incubators. SHAP elucidates machine learning model predictions by assigning importance values to each feature based on game theory, ensuring consistent and locally accurate feature attributions. In this study, we employed a gradient-based approximation of SHAP values \cite{49_jethani2021fastshap}.

To maintain analytical consistency, we applied SHAP to the best-fit model for each ecosystem. Our objective was to assess how the model behaves when classifying previously unseen graduated projects. Therefore, we applied SHAP to the 20\% data split of each different sets (ASF, EF, OF and G). For this analysis, we excluded all retired projects to focus on graduated ones. For each graduated project within the 20\% ASF project data split, we computed the individual feature importance. Next, we aggregated these individual feature importance values to derive a global interpretation for the ASFI. This methodology was similarly applied to obtain the global feature importance for OF, EF and GitHub. We then performed a processing step to determine the relative importance of each feature within an incubator. 

To identify the relative importance of each feature, we normalized both positive and negative SHAP values such that their magnitudes collectively sum to 100. First, we computed the total magnitude of all positive SHAP values (\texttt{total positive magnitude}) and all negative SHAP values \texttt{(total negative magnitude)} separately. We then divided each positive value by \texttt{total positive magnitude} and each negative value by \texttt{total negative magnitude}, and multiplied the results by 100. This yielded a final set of values on a scale from $-100$ to $+100$, reflecting the relative contribution and direction of each feature.

\subsection{OSS-ProF: OSS Project-Foundation Router}

To produce a single generalized starting point for tracking any project's sustainability,
we developed the OSS Project-Foundation Router (OSS-ProF), a model that classifies an unseen project to a software foundation (ASFI, ESF, OFI, or GitHub/none), with which the project most closely aligns, based on its sociotechnical features.

\begin{figure}[]
\includegraphics[width = \linewidth, height=6.6cm]{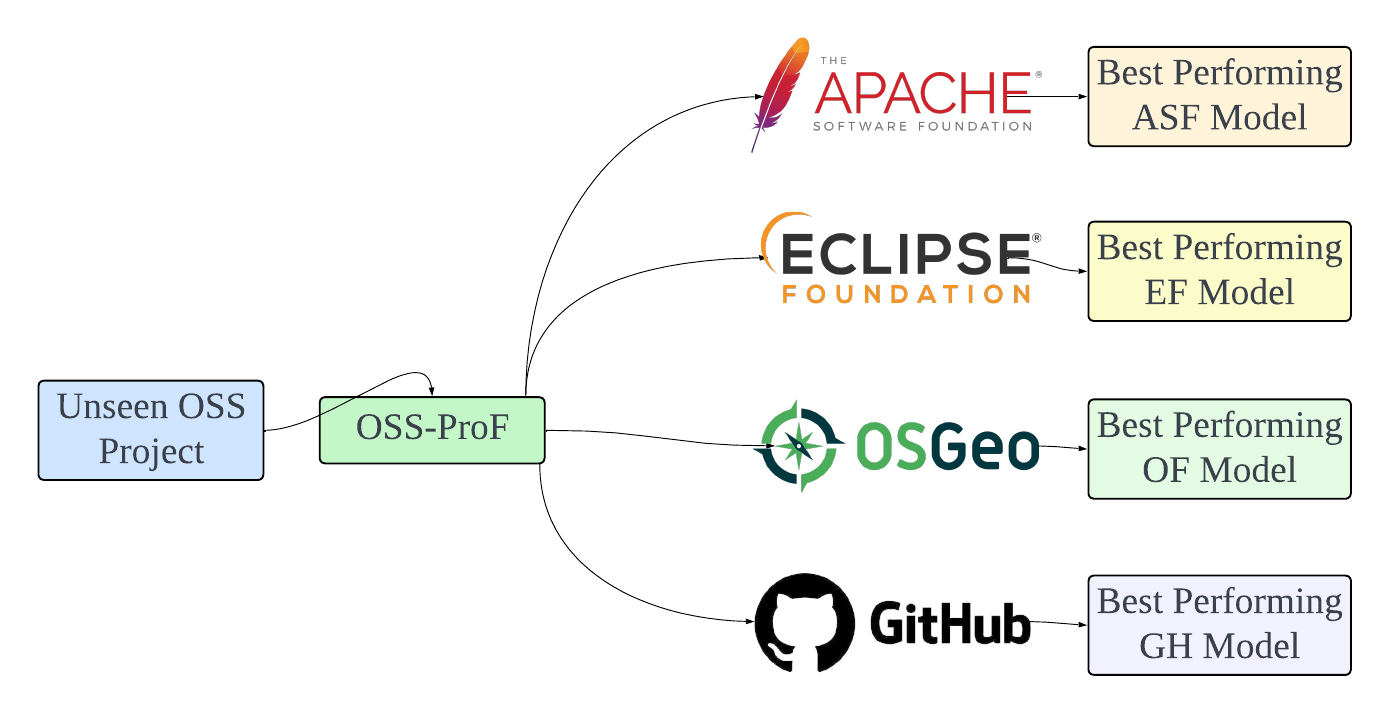}
\caption{Overview of the OSS-ProF two-layered pipeline. The first layer (OSS Project–Foundation Sorter) classifies each unseen project into the most analogous foundation—Apache, Eclipse, OSGeo, or GitHub—based on its socio-technical features. The second layer then applies the corresponding foundation-specific deep learning model to forecast the project’s sustainability outcome. This two-step pipeline enables context-aware sustainability prediction across heterogeneous OSS ecosystems.}
\label{project_router}

\end{figure}

To answer RQ2, we use OSS-ProF to decide which foundation model is most appropriate for the given project, and then use that model to predict project sustainability. Thus, we are using a 2-step AI pipeline, as shown in Figure \ref{project_router} consisting of a classifier upstream that chooses the deep learning model downstream.  
OSS-ProF was trained using the same sociotechnical feature set and modeling pipeline as the sustainability models, while its target variable represents the foundation label, Apache, Eclipse, OSGeo, or GitHub, rather than sustainability outcome.

\textbf{Training Protocol.} The OSS-ProF router and foundation-specific models are trained independently with no shared weights. We maintain consistent train/test splits: projects held out for testing are unseen by both the router and all downstream sustainability models during training. This ensures no data leakage in the combined pipeline evaluation (Table~6).






\section{Results}

\subsection{$\textbf{\textit{RQ}}_{\textcolor{red}{\textit{1}}}$: Does a unified model of sustainability based on sociotechnical network features exist between the ASF, EF, and OF foundations?}

We trained models on sociotechnical features in different combinations of ecosystems. 


\textbf{General Trials (Intra-Foundation).}  
When models were trained and evaluated within the same foundation, we observed consistently high performance across all three architectures. For context, intra-foundation trials used 5-fold stratified cross-validation with the median score being picked. For example, for ASFI, Bi-LSTM achieved an F1-score of 95.60\%, with Bi-DLSTM close behind at 92.66\%. Eclipse also saw strong results, with Bi-LSTM and Bi-DLSTM reaching 90.21\% and 90.01\% F1-scores, respectively. OFI similarly yielded robust outcomes, especially with Bi-DLSTM achieving an F1-score of 89.29\%. These results indicate that sustainability modeling is highly effective when constrained to the specific characteristics of a given foundation.

\textbf{Transfer Trials (Cross-Foundation).}  
In contrast, performance generally declined in cross-foundation transfer scenarios. When models trained on ASFI were applied to EFI, F1-scores dropped—for instance, Bi-LSTM scored 70.55\%, and Transformer dropped to 55.01\%. EFI-to-ASFI showed slightly better generalization, with Bi-LSTM and Bi-DLSTM achieving F1-scores around 73–74\%. Transfer from ASFI or EFI to OFI also resulted in performance degradation (Bi-LSTM: ~73–76\%, Bi-DLSTM: ~67–71\%). These findings highlight the challenge of transferring sustainability signals across incubators, likely due to differing governance models, project lifecycles, and evaluation criteria.

\textbf{Extrapolation to and from GitHub.}  
When predicting GitHub project sustainability, training directly on GitHub produced the best results; Bi-DLSTM achieved an F1-score of 97.62\%, and Bi-LSTM reached 87.46\%. Interestingly, training on ASFI or EFI also generalized well for GH projects, with Transformer achieving up to 93.52\% F1 in the ASFI case. However, training on OFI and testing on GitHub (O to G) resulted in lower performance (Bi-LSTM: 57.14\%, Bi-DLSTM: 55.10\%), suggesting limited representational overlap.

In the reverse direction, models trained on GitHub data generalized poorly to foundation projects. For example, GH-to-EFI and GH-to-ASFI produced low F1-scores across all models, with Bi-DLSTM performing worst (30–40\% F1). This asymmetry suggests that while GitHub projects may mimic sustainability patterns from structured foundations during early development, the reverse does not hold—likely due to GitHub’s lack of formal incubation, governance, and evaluation mechanisms.


\noindent
\fbox{%
  \parbox{\linewidth}{%
    \textbf{$\textbf{\textit{RQ}}_{\textcolor{red}{\textit{1}}}$ Findings:}  
Sustainability models trained on sociotechnical networks are highly effective when evaluated within the same foundation. However, cross-foundation generalization remains limited, especially when transferring between structurally different ecosystems. While GitHub projects can be predicted reasonably well using models trained on ASFI or EFI, the reverse is not true; models trained on GitHub fail to generalize back to structured foundations. Bi-LSTM models show the strongest performance overall, particularly in capturing foundation-specific sustainability patterns.
  }%
}

To assess result stability, Table~\ref{tab:std_dev_results} in the appendix \ref{cv_variance} reports the standard 
deviation of F1-scores across fifteen independent runs per trial. Intra-foundation trials demonstrate low variance for ASF ($\sigma = 0.0095$) and EF ($\sigma = 0.0134$), confirming consistent convergence. However, trials involving OSGeo and GitHub exhibit notably higher variance (up to 
$\sigma = 0.1367$), attributable to their smaller sample sizes. Accordingly, median F1-scores for these ecosystems should be interpreted with caution.



\begin{table*}[ht]
\centering
\caption{Model performance on \textit{intra-foundation} trials, where each model is trained on 80\% of its foundation’s data and tested on the remaining 20\% with 5-fold CV. Weighted F1 is reported.}
\label{table:SoloFoundationResults}
\resizebox{\textwidth}{!}{\renewcommand{\arraystretch}{0.85}
\begin{tabular}{@{}ccccccccccc@{}}
\toprule
\multicolumn{2}{c}{\textbf{Combination}} 
& \multicolumn{3}{c}{\textbf{Bi-LSTM}} 
& \multicolumn{3}{c}{\textbf{Transformer}} 
& \multicolumn{3}{c}{\textbf{Bi-DLSTM}} \\
\cmidrule(lr){1-2} \cmidrule(lr){3-5} \cmidrule(lr){6-8} \cmidrule(lr){9-11}
 {Train} & {Test}
 & {Precision} & {Recall} & {F1 Score} 
 & {Precision} & {Recall} & {F1 Score} 
 & {Precision} & {Recall} & {F1 Score} \\
\midrule
ASFI & ASFI & 95.96 & 95.51 & \textbf{95.60} & 91.30 & 87.18 & 87.59 & 93.49 & 92.31 & 92.66 \\
EFI & EFI & 91.18 & 89.37 & \textbf{90.21} & 66.11 & 90.32 & 85.73 & 91.45 & 88.77 & 90.01 \\
OFI & OFI & 84.72 & 75.00 & 73.33 & 84.72 & 75.00 & 74.44 & 95.14 & 87.50 & \textbf{89.29} \\
GH & GH & 100.00 & 83.33 & 87.46 & 94.44 & 91.67 & 91.11 & 100.00 & 95.83 & \textbf{97.62} \\
\bottomrule
\end{tabular}}
\end{table*}

\begin{table*}[ht]
\centering
\caption{Performance of the models on \textit{cross-foundation} trials. Each model is trained on the full dataset (100\%) of one foundation and evaluated on the complete dataset of another foundation. Weighted F1 is reported.}
\resizebox{\textwidth}{!}{\renewcommand{\arraystretch}{0.85}
\label{table:CrossIncubatorResults}
\begin{tabular}{@{}ccccccccccc@{}}
\toprule
\multicolumn{2}{c}{\textbf{Combination}} 
& \multicolumn{3}{c}{\textbf{Bi-LSTM}} 
& \multicolumn{3}{c}{\textbf{Transformer}} 
& \multicolumn{3}{c}{\textbf{Bi-DLSTM}} \\
\cmidrule(lr){1-2} \cmidrule(lr){3-5} \cmidrule(lr){6-8} \cmidrule(lr){9-11}
 {Train} & {Test}
 & {Precision} & {Recall} & {F1 Score} 
 & {Precision} & {Recall} & {F1 Score} 
 & {Precision} & {Recall} & {F1 Score} \\
\midrule
ASFI & EFI & 89.56 & 63.31 & \textbf{70.55} & 85.44 & 46.04 & 55.01 & 88.32 & 55.40 & 63.78 \\
ASFI & OFI & 76.64 & 73.33 & 73.64 & 79.50 & 75.00 & \textbf{75.58} & 74.48 & 68.33 & 67.94 \\
\addlinespace
EFI & ASFI & 83.75 & 79.21 & \textbf{73.73} & 60.76 & 77.95 & 68.29 & 83.51 & 79.09 & 71.77 \\
EFI & OFI & 76.49 & 73.33 & \textbf{69.27} & 42.25 & 65.00 & 51.21 & 71.97 & 70.83 & 67.54 \\
\addlinespace
OFI & ASFI & 75.54 & 77.57 & 75.63 & 85.92 & 85.80 & \textbf{84.51} & 77.04 & 79.72 & 77.02 \\
OFI & EFI & 88.08 & 75.54 & 79.66 & 88.45 & 85.85 & \textbf{86.85} & 85.93 & 80.34 & 82.50 \\

\bottomrule
\end{tabular}}
\end{table*}

\begin{table*}[ht]
\centering
\caption{Extrapolation to and from GitHub: performance of the models involving the GitHub ecosystem.
The first three combinations represent transfer from foundation-trained models (ASF, EF, OF) to GitHub projects, while the last three correspond to models trained on GitHub and tested on foundation datasets. GitHub is treated separately as it represents an open-source ecosystem rather than a structured foundation.}
\label{table:GResults}
\resizebox{\textwidth}{!}{\renewcommand{\arraystretch}{0.85}

\begin{tabular}{@{}ccccccccccc@{}}

\toprule
\multicolumn{2}{c}{\textbf{Combination}} 
& \multicolumn{3}{c}{\textbf{Bi-LSTM}} 
& \multicolumn{3}{c}{\textbf{Transformer}} 
& \multicolumn{3}{c}{\textbf{Bi-DLSTM}} \\
\cmidrule(lr){1-2} \cmidrule(lr){3-5} \cmidrule(lr){6-8} \cmidrule(lr){9-11}
 {Train} & {Test}
 & {Precision} & {Recall} & {F1 Score} 
 & {Precision} & {Recall} & {F1 Score} 
 & {Precision} & {Recall} & {F1 Score} \\
\midrule
ASFI & GH & 89.16 & 87.30 & 85.99 & 94.19 & 93.65 & \textbf{93.52} & 92.86 & 92.06 & 91.70 \\
EFI & GH & 82.74 & 79.37 & 79.60 & 66.82 & 74.60 & 68.22 & 83.97 & 76.19 & \textbf{77.69} \\
OFI & GH & 64.35 & 57.14 & 57.14 & 87.51 & 87.30 & \textbf{87.25} & 67.86 & 53.97 & 55.10 \\
\addlinespace
GH & ASFI & 84.18 & 63.88 & 66.75 & 87.67 & 75.41 & \textbf{77.42} & 82.73 & 40.18 & 38.56 \\
GH & EFI & 84.87 & 47.96 & \textbf{57.04} & 85.20 & 38.13 & 46.32 & 83.27 & 30.70 & 37.42 \\
GH & OFI & 71.75 & 70.00 & 70.42 & 74.74 & 71.67 & \textbf{70.73} & 63.39 & 57.50 & 57.77 \\

\bottomrule
\end{tabular}}
\end{table*}

\subsection{$\textbf{\textit{RQ}}_{\textcolor{red}{\textit{2}}}$: Based on sociotechnical network features, can we identify the software foundation to which an open-source project is most analogous, and, once routed, can the corresponding foundation-specific sustainability model effectively forecast its sustainability outcome?}



To address RQ2, we developed OSS-ProF, which was trained on the proposed models while utilizing the sociotechnical data where target variable represents the foundation labels. The Bi-LSTM model yielded the best overall performance, achieving a precision of 96.21\%, recall of 91.05\%, and an F1-score of 93.63\%. The Transformer model followed closely, with precision of 92.36\%, recall of 86.15\%, and an F1-score of 89.25\%. The Bi-DLSTM model achieved a precision of 85.92\%, recall of 79.22\%, and an F1-score of 82.57\%. Across all models, the OSS-ProF achieved reasonable performance, demonstrating that sociotechnical features contain sufficient signal to distinguish between foundations with distinct governance and collaboration structures. 

To assess the practical utility of this routing stage, we integrated OSS-ProF with the corresponding foundation-specific sustainability models (see Table~\ref{triager_perf}). The combined “OSS-ProF + Best Model” configuration markedly improved sustainability forecasting performance across all ecosystems. For Apache and Eclipse, the routed models achieved F1-scores of 96.20\% and 96.15\%, respectively, representing a consistent gain over their standalone foundation-specific forecasters. OSGeo and GitHub exhibited strong results with 92.83\% and 95.76\% F1, though these values should be interpreted cautiously given their comparatively small sample sizes. Though 5-fold cross-validation was used, the limited number of projects in OFI and GH enabled the routing architecture to correctly map a higher proportion of samples, enabling stronger downstream performance from the sustainability models. Overall, these results confirm that accurate project–foundation routing substantially enhances sustainability inference by ensuring that each project is evaluated by the most contextually appropriate, foundation-specific predictive framework.

\begin{table}[h!]
\centering
\caption{Comparison of sustainability prediction performance between foundation-specific best models and the combined OSS-ProF + Foundation-specific Best Model setup. In the combined setting, projects are first routed through the OSS-ProF, which determines the most suitable foundation for each project. The corresponding foundation-specific best model is then applied to predict the project’s sustainability outcome.}
\label{triager_perf}
\begin{tabular}{lcccccc}
\hline
\multirow{2}{*}{\textbf{Foundation}} &
\multicolumn{3}{c}{\textbf{Best Model}} &
\multicolumn{3}{c}{\shortstack{\textbf{with OSS-ProF}\\\textbf{+ Best Model Classifier}}} \\ \cmidrule(lr){2-4} \cmidrule(lr){5-7}

 & \textbf{P} & \textbf{R} & \textbf{F1} & \textbf{P} & \textbf{R} & \textbf{F1} \\ \hline
ASFI & 95.96 & 95.51 & 95.60 & 97.44 & 95.00 & 96.20 \\
EFI  & 91.18 & 89.37 & 90.21 & 100.00 & 92.59 & 96.15 \\
OFI  & 95.14 & 87.50 & 89.29 & 96.88 & 90.00 & 92.83 \\
GH   & 100.00 & 95.83 & 97.62 & 96.88 & 95.00 & 95.76 \\ \hline
\end{tabular}
\end{table}

\noindent
\fbox{%
  \parbox{\linewidth}{%
    \textbf{$\textbf{RQ}_{\textcolor{red}{2}}$ Findings:}  Yes, based on a project’s sociotechnical features, we can develop an OSS Project–Foundation Router (OSS-ProF) that predicts which open-source ecosystem a project most closely aligns with. OSS-ProF, when paired with foundation-specific sustainability forecasters, substantially improves predictive performance across all foundations, confirming the effectiveness of the two-step routing–forecasting pipeline.
  }%
}

\subsection{$\textbf{\textit{RQ}}_{\textcolor{red}{\textit{3}}}$: What is the relationship between models of OSS project success and OSS project sustainability? Can our sustainability models be used with projects outside of foundations?}

Sustainability and success are often used interchangeably in open-source software (OSS) research, yet they may represent distinct concepts. \textit{Sustainability}, as defined within software foundations such as ASF and EF, typically refers to a project's ability to build a long-term, diverse, and self-governing community capable of producing regular, well-governed software releases. In contrast, \textit{success}—particularly in decentralized platforms like GitHub—is often tied to popularity metrics, contributor activity, or short-term momentum, as characterized in the study by Joblin et al.~\cite{6_joblin2022successful}. This raises the question: to what extent are sustainability and success overlapping constructs, and can models trained to detect one accurately predict the other?

To investigate this, we conducted two sets of transfer experiments. In the first, we trained models on foundation-labeled sustainability data (e.g., graduated vs. retired in ASF/EF/OSGeo) and tested them on GitHub projects labeled as successful or unsuccessful. In the second, we reversed this: models were trained on GitHub success labels and evaluated on foundation datasets labeled for sustainability.

The results reveal a nuanced relationship. Models trained on sustainability data—particularly from Apache and Eclipse—generalized reasonably well to GitHub success prediction. For example, the Transformer model trained on Apache (A $\rightarrow$ G) achieved an F1-score of 93.52\%, while Bi-LSTM achieved 85.99\%. These high scores suggest that sustainability signals in structured foundations often align with early markers of success in GitHub projects, such as sustained contributions, governance structures, and active collaboration. In other words, successful GitHub projects tend to mirror sustainable ones in their formative behaviors.

However, the reverse direction—training on GitHub and testing on foundations—produced significantly weaker results. For example, models trained on GitHub and tested on Eclipse (G $\rightarrow$ E) yielded F1-scores as low as 37.42\% (Bi-DLSTM), and similar degradation was seen for G $\rightarrow$ A. This indicates that success signals learned from GitHub do not generalize well to the more rigorous and multifaceted sustainability standards enforced in incubator foundations. GitHub success often lacks institutional markers like governance maturity, licensing compliance, and community diversity, which are critical for foundation-based sustainability outcomes.

This asymmetry highlights an important insight: while sustainability appears to subsume many early signals of success, success alone is insufficient to indicate long-term sustainability. High contributor activity or growth does not guarantee that a project will achieve graduation, especially if it lacks governance structure or inclusive collaboration. \newline

\noindent
\fbox{%
  \parbox{\linewidth}{%
    \textbf{$\textbf{\textit{RQ}}_{\textcolor{red}{\textit{3}}}$ Findings:}  
Models trained on sustainability-labeled data from structured foundations (e.g., ASF, EF) can reasonably predict GitHub project success, indicating alignment between sustainability practices and early success signals. However, models trained on GitHub-labeled success data perform poorly when applied to sustainability prediction in foundations. This suggests that while sustainability often includes early indicators of success, the notion of success in GitHub projects does not fully capture the diverse manifestations of sustainability across foundation ecosystems.
  }%
}



\subsection{$\textbf{\textit{RQ}}_{\textcolor{red}{\textit{4}}}$: Which sociotechnical features are most predictive of sustainability in the best-fit models for different foundations and ecosystems?}

\begin{figure*}[!ht]
\includegraphics[width =10cm, height=10cm]
{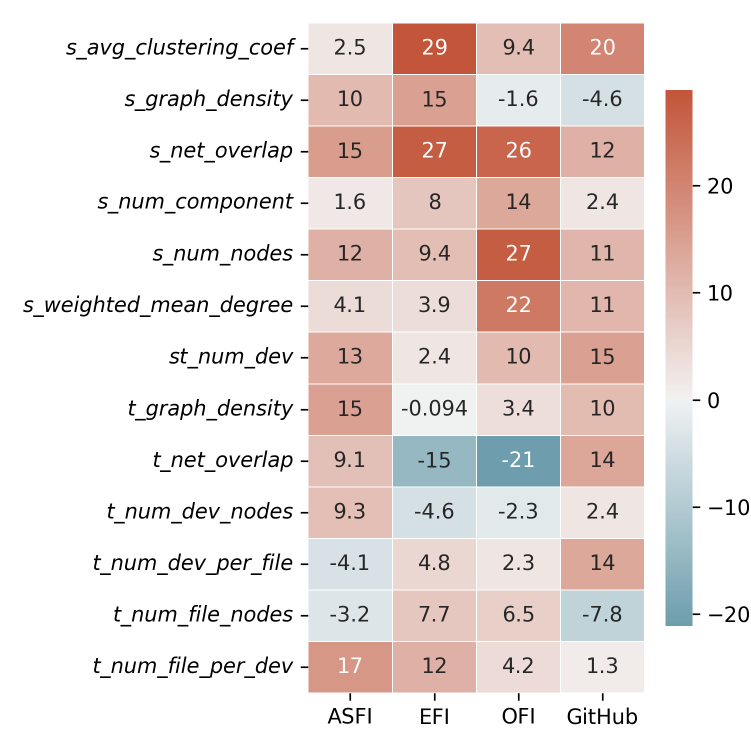}
\centering
\caption{Feature Importance of Models for Predicting Sustainability Outcomes Across ASFI, EFI, OFI, and GH}
\label{feature_imp}
\end{figure*}

To investigate which sociotechnical features are most predictive of sustainability outcomes across different foundations, we computed feature importances using SHAP-based explainability methods applied to the best-performing models for each ecosystem. Figure~\ref{feature_imp} presents the normalized importance values for a shared set of features across the ASFI, EFI, OFI, and GH ecosystems. 


\textbf{Common Predictive Signals.}  
Several features show consistently high importance across multiple ecosystems, indicating a set of generalizable sustainability signals. For example, \texttt{s\_net\_overlap}, which measures how connected the structural network remains over time, is highly predictive in ASFI (15), EFI (27), OFI (26), and GitHub (12). This suggests that sustained, overlapping collaboration patterns are a strong indicator of sustainability, regardless of the foundation. Similarly, \texttt{s\_num\_nodes}, the number of active contributors, has high values across ASFI (12), EFI (9.4), OFI (27), and GitHub (11), reaffirming the importance of contributor engagement. \texttt{s\_graph\_density} and \texttt{t\_graph\_density} also show relevance, indicating that both structural and temporal cohesion play a significant role.

\textbf{Foundation-Specific Drivers.}  
While commonalities exist, several features show pronounced importance only in specific ecosystems. E.g., \texttt{s\_avg\_\allowbreak clustering\_\allowbreak coef} is a key factor in EFI (29) and GitHub (20), but less so in ASF and OFI. This may reflect differences in how tightly-knit contributor subgroups form and persist in these environments. Likewise, \texttt{s\_\allowbreak weighted\_\allowbreak mean\_\allowbreak degree} is notably influential in OFI (22), suggesting that in OSGeo, the sustainability of projects may depend more on high centrality or influence of individual contributors. In contrast, \texttt{st\_num\_dev}, the combined structural and temporal measure of developer engagement is predictive in ASF (13) and GitHub (15), potentially highlighting the foundational emphasis on long-term contributor retention.

\textbf{Temporal Dynamics and Variance.}  
Temporal features reveal further differentiation. The feature \texttt{t\_num\_file\_per\_dev}, representing how much file responsibility is shared among contributors, is important in ASF (17) and EFI (12), indicating that workload balance may correlate with sustainability in structured incubator settings. On the other hand, GitHub exhibits higher importance in \texttt{t\_num\_dev\_per\_file} (14) and \texttt{t\_net\_overlap} (14), implying that file-level collaboration and persistence of developer involvement are more critical in informal OSS environments. Interestingly, OFI assigns modest importance across a broader range of features, suggesting that sustainability in OSGeo projects may stem from a more distributed set of signals rather than a few dominant factors.

\textbf{Interpreting Low or Negative Importance.}  
Some features exhibit low or even negative importance in specific foundations. E.g., \texttt{t\_net\_overlap} is negatively associated with sustainability in EFI (-15) and OFI (-21), which may reflect differences in how temporal continuity is valued or measured. Similarly, \texttt{t\_num\_dev\_nodes} shows negative influence in EFI (-4.6) and OFI (-2.3), indicating that more fragmented developer-file interactions might hinder sustainability in these ecosystems. GitHub, however, has positive weights to many of these features, reinforcing that OSS projects outside structured foundations operate under different dynamics.

\textbf{Implications.}  
While a shared modeling framework can capture general patterns of sustainability, the drivers behind sustainability are not uniform  across foundations. Governance models, contributor onboarding practices, and community expectations can all shape which sociotechnical signals are most salient. Therefore, interpreting model outputs must be done in the context of the ecosystem, and sustainability interventions should be tailored accordingly.

\noindent
\fbox{%
  \parbox{\linewidth}{%
  \textbf{$\textit{RQ}_{\textcolor{red}{\textit{4}}}$ Findings:}  While some sociotechnical features, such as social net overlap (\texttt{s\_net\_overlap}) and social contributor count (\texttt{s\_num\_nodes}), positively correlate with sustainability across ecosystems, others are foundation-specific. E.g., the clustering coefficient and mean degree are more predictive in EFI and OFI, while GitHub projects emphasize temporal collaboration continuity. This suggests that sustainability models should incorporate both shared and ecosystem-specific features to improve generalizability and actionable insight.
  }%
}



\section{Case Studies}
\begin{figure*}[]
  \centering
  \begin{minipage}[]{0.45\textwidth}
    \centering
    \includegraphics[width=\linewidth]{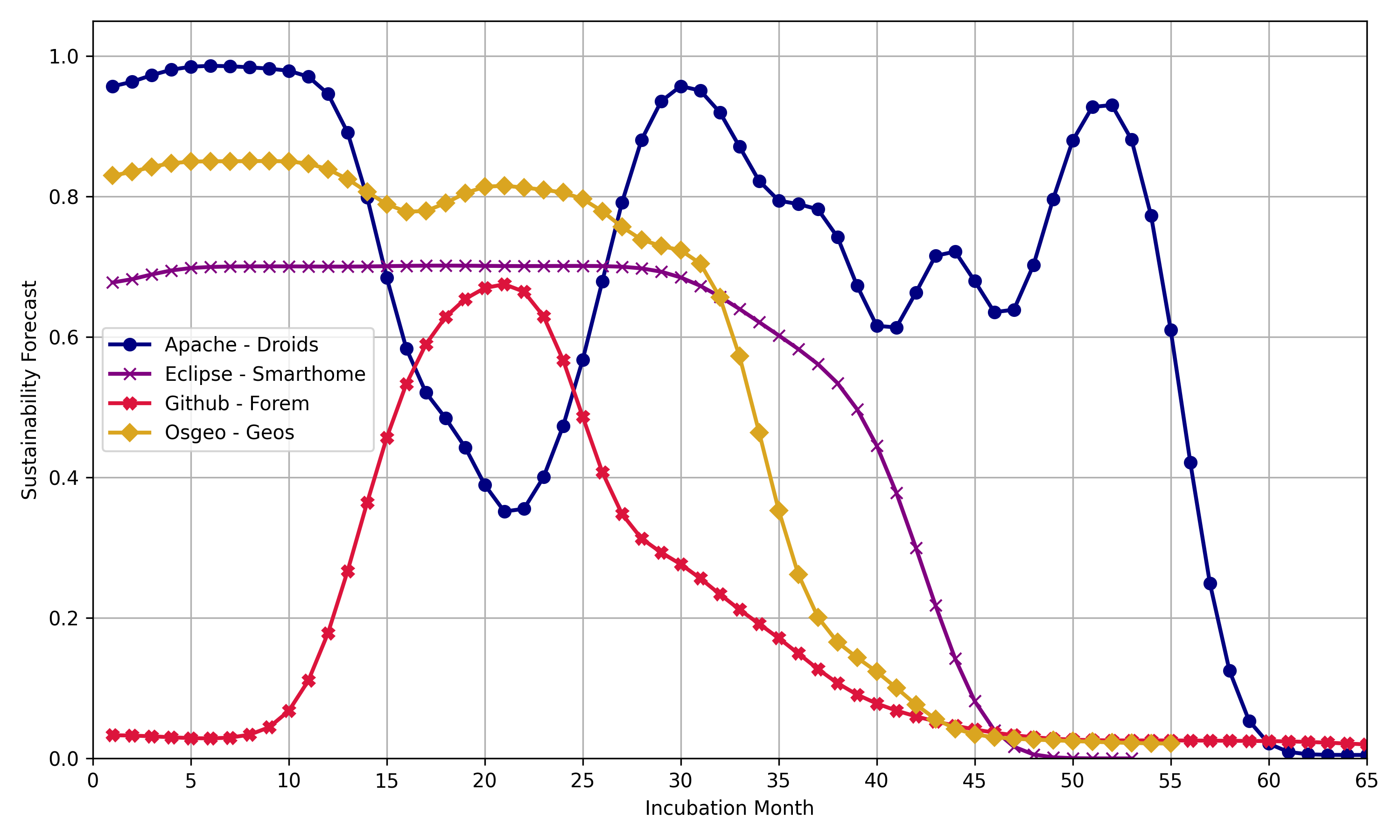}
    \captionof{figure}{Sustainability Forecasts for Four Failed Projects Across Different Ecosystems:
    1) \emph{Droids} — Apache, 2) \emph{SmartHome} — EF, 3) \emph{Forem} — GitHub, and 4) \emph{Geos} — OSGeo}
    \label{case_study_forecast}
  \end{minipage}%
  \hspace{0.02\textwidth}
  \begin{minipage}[]{0.45\textwidth}
    \centering
   
    \setlength{\tabcolsep}{2pt}   
    \renewcommand{\arraystretch}{1.05}
    \scriptsize
    \resizebox{\linewidth}{!}{%
      \begin{tabular}{@{}lrrrr@{}}
        \toprule
        \textbf{Feature} & \textbf{Droids (ASF)} & \textbf{SmartHome (EF)} & \textbf{Geos (OF)} & \textbf{Forem (GH)} \\
        \midrule
        s\_num\_nodes              & -0.23 & -1.84 & -0.16 & -0.25 \\
        s\_weighted\_mean\_degree  & -0.76 & -3.58 & -0.89 & -0.17 \\
        s\_num\_component          & -1.46 & -0.84 & -0.92 & -0.29 \\
        s\_avg\_clustering\_coef   & -0.34 & -3.84 &  0.32 & -0.22 \\
        s\_graph\_density          & -0.69 & -0.36 & -0.53 & N/A   \\
        s\_net\_overlap            & -0.23 & -0.25 & N/A   & -0.20 \\
        \midrule
        st\_num\_dev               & -0.76 & -0.92 & -0.84 & -0.07 \\
        \midrule
        t\_num\_dev\_nodes         & -0.61 & N/A   & N/A   & -2.40 \\
        t\_num\_file\_nodes        & N/A   &  2.84 &  0.10 & N/A   \\
        t\_num\_dev\_per\_file     & N/A   & -0.05 &  0.04 &  0.07 \\
        t\_num\_file\_per\_dev     & -0.85 &  0.20 & -0.18 & -0.25 \\
        t\_graph\_density          & -0.61 & N/A   & -0.30 &  0.51 \\
        t\_net\_overlap            & -0.86 & N/A   & N/A   & -0.04 \\
        \bottomrule
      \end{tabular}%
    }
     \captionsetup{type=table}
    \caption{Feature Deviations During Downturn Periods Across the Four Failed Projects}
    \label{tab:feature-impact}
  \end{minipage}
\end{figure*}
To validate the interpretability and robustness of our proposed framework, we conducted a series of case studies across four representative failed projects—\texttt{Droids} (Apache), \texttt{SmartHome} (Eclipse), \texttt{Geos} (OSGeo), and \texttt{Forem} (GitHub). The objective was to cross-check whether our sustainability forecasting approach can (a) correctly capture the observed sustainability trajectory of individual projects over time, and (b) identify, through feature-importance analysis, which sociotechnical attributes were most affected during periods of downturn. This examination  also allowed us to assess whether the factors contributing to project decline are consistent with the global feature-importance trends reported earlier.

Each project was evaluated using the best-performing foundation-specific model. Figure~\ref{case_study_forecast} illustrates the monthly sustainability forecast probabilities for each project, based on the sociotechnical traces of preceding months. The forecasts reveal that all four projects exhibited a gradual decline leading to eventual inactivity—\texttt{Droids} around month~61, \texttt{SmartHome} near month~48, \texttt{Forem} by month~45, and \texttt{Geos} at month~46 indicating that the models successfully tracked temporal degradation.

\noindent\textbf{Downturn Feature Analysis.}
To examine which sociotechnical signals shifted most during decline, we defined a \textit{downturn interval} spanning six months before and after each project’s observed inflection point (e.g., \texttt{Droids}: months~49–61; \texttt{SmartHome}:~24–36; \texttt{Forem}:~14–26; \texttt{Geos}:~26–38). For each feature, we computed the mean value during the pre-downturn period as a baseline and compared it to the post-downturn values to obtain deviation scores. Negative deviations correspond to decreases in features previously shown to have a positive effect on sustainability—signaling sociotechnical weakening—whereas positive deviations indicate recovery or stability.

Table~\ref{tab:feature-impact} summarizes the maximum post-downturn deviations for the positively impactful features across projects. In every case, features that contributed most strongly to sustainability in our global models—such as developer overlap (\texttt{s\_net\_overlap}), contributor engagement (\texttt{s\_num\_nodes}), and workload balance \newline (\texttt{t\_num\_file\_per\_dev})—declined sharply during downturn periods. This alignment between project-level feature deterioration and the global importance rankings reported earlier provides an empirical consistency check for our framework, that he same sociotechnical mechanisms driving sustainability at ecosystem level are reflected in the micro-level dynamics preceding project failures.

\noindent
\fbox{%
  \parbox{\linewidth}{%
    \textbf{Case Study Findings:} 
Across ecosystems, the case studies validate that our forecasting models capture the sustainability trajectory of projects over time and pinpoint deteriorating sociotechnical factors. The observed downturn patterns are consistent with the global feature-importance results, reinforcing the framework's explanatory ability.
  }%
}





\section{Discussion}
\subsection{Universal Sustainability Prediction}
A central contribution of this work is the OSS-ProF framework, which addresses a fundamental challenge in cross-ecosystem sustainability modeling: institutional heterogeneity. Contingency theory in organizational research posits that there is no universally optimal structure or process; rather, organizational effectiveness depends on achieving fit between environmental context and internal mechanisms~\cite{contingency_ref_1, contingency_ref_2}. In OSS ecosystems, this manifests as divergent governance models, communication norms, IP policies, and community-building practices across foundations; what works in Apache's consensus-driven, mailing-list-centric incubation may not translate to Eclipse's corporate-backed PMC oversight or OSGeo's domain-expert review panels. Our RQ1 cross-foundation transfer experiments empirically confirm this: models trained on one foundation's sustainability patterns fail when applied to another (F1 scores drop 15-40\%), demonstrating that sustainability is not a universal construct but rather context-dependent. Rather than forcing all projects through a single sustainability classifier—an approach our results show to be ineffective—OSS-ProF operationalizes contingency theory as a mixture-of-contexts model through a two-stage prediction pipeline (Figure~\ref{project_router}). When an unseen OSS project is presented to the system, its sociotechnical features are first analyzed by the OSS-ProF router, a lightweight classifier trained to identify which software foundation ecosystem the project most closely resembles based on its collaboration patterns, governance signals, and technical dynamics. Once classified (as Apache-like, Eclipse-like, OSGeo-like, or GitHub-like), the project is routed to the corresponding foundation-specific sustainability model, which has been optimized to capture the unique sociotechnical patterns of that ecosystem. This architecture aligns with sociotechnical fit theory—the principle that alignment between work patterns and organizational structures drives project outcomes. The routing stage learns to recognize governance context from sociotechnical signatures, enabling downstream models to leverage context-specific sustainability signals rather than diluting predictive power through forced generalization across incompatible institutional environments.

This routing strategy yields substantial improvements over both single-foundation models applied universally and cross-foundation transfer approaches. As shown in Table 5, the combined "OSS-ProF + Best Model" achieves F1-scores of 96.20\% for Apache-like projects, 96.15\% for Eclipse-like projects, 92.83\% for OSGeo-like projects, and 95.76\% for GitHub projects. These results represent consistent gains of 3-8\% over naive cross-foundation application, demonstrating that accurate ecosystem identification is critical for reliable sustainability forecasting. For practitioners, this has direct implications: rather than guessing which foundation's sustainability criteria apply to a new project, OSS-ProF provides data-driven guidance on where a project "fits" and which evaluation standards are most appropriate. This is particularly valuable for projects considering joining a foundation, for foundation mentors triaging incoming proposals, or for funding bodies assessing long-term project viability across diverse governance contexts.


A critical question for the generalizability of our findings is: what portion of the OSS foundation landscape do Apache, Eclipse, and OSGeo represent? While no universal taxonomy exists for categorizing all foundation characteristics, we can systematically map the governance, operational, and community attributes captured by our study (Table \ref{tab:foundation_coverage} in Appendix \ref{foundation_coverage}).

\subsection{Characteristics Covered by Our Study}
Each foundation in our sample contributes distinct and complementary characteristics to our coverage space. The Apache Software Foundation brings meritocratic governance structures \cite{mockus2007large}, community-driven decision-making through consensus voting, and mandatory mailing list communication—representing the archetype of volunteer-driven, transparency-focused OSS governance \cite{jergensen2011onion}. Apache's donation-based funding model and strict adherence to the "Apache Way" during incubation exemplify decentralized, principle-driven project oversight at scale (350+ projects across diverse domains). The Eclipse Foundation contributes a fundamentally different model: corporate-backed governance with vendor-neutral oversight, rigorous intellectual property review processes, and Project Management Committee (PMC) structures that balance corporate and community interests \cite{gaughan2009examination}. Eclipse's membership-fee funding model and enterprise-focused project portfolio represent the corporate-OSS hybrid that has become increasingly prevalent in modern software ecosystems. OSGeo adds domain-specific governance through expert review panels, research-oriented community structures with strong academic ties, and flexible multi-tier incubation pathways tailored to geospatial technology needs \cite{steiniger20132012, ramseystate}. OSGeo's grant-and-conference funding model and smaller, specialized project scale represent niche-domain foundations serving specific technical communities.

Despite this breadth, our study does not capture all foundation archetypes. Notably absent are single-vendor controlled foundations (e.g., Google's Android Open Source Project, Meta's React), where corporate ownership fundamentally alters sustainability dynamics and graduation criteria. We also do not include cloud-native specialized foundations (Cloud Native Computing Foundation), scientific computing foundations (NumFOCUS), or language-specific foundations (Python Software Foundation, Ruby Central), each of which may impose domain-specific sustainability requirements beyond our captured governance patterns. Additionally, infrastructure-focused foundations like the Linux Foundation—which emphasizes kernel and systems-level projects with distinct technical collaboration patterns—remain outside our sample. Finally, hybrid proprietary-OSS models and dual-licensing strategies represent governance approaches we do not directly model.



The diversity of characteristics captured, spanning volunteer-to-corporate contributor bases, donation-to-membership funding, consensus-to-hierarchical decision-making, and generalist-to-specialist project domains—suggests that our framework captures sustainability signals across major governance archetypes. This interpretation is further supported by our successful generalization to GitHub projects (RQ3), which operate outside any foundation structure entirely. While we acknowledge gaps in coverage, particularly around single-vendor and cloud-native models, the triangulation across three structurally distinct foundations provides a robust empirical basis for understanding multi-ecosystem sustainability prediction. Future validation on Linux Foundation, CNCF, or NumFOCUS projects would clarify whether uncovered characteristics introduce fundamentally different sustainability dynamics or whether diminishing returns occur as additional foundations are incorporated.

\subsection{Practitioner Takeaways}
For individual projects or engineering leaders, the lesson is to treat “project sustainability” in practice as context-specific, not universal. Our models trained within a governance context (e.g., ASFI vs. EFI vs. OFI vs. general GitHub) perform best when evaluated in that same context. This means one will get more reliable triage and risk flags by first routing projects to their closest governance context (foundation-like vs. GitHub) and then applying context-specific predictors rather than a single global model. The router+predictor can be used to segment watch-lists into (a) projects needing immediate support (maintainer outreach, backlog triage, funding), (b) projects stable but trending worse (preemptive mentorship, governance tune-ups), and (c) projects suitable for deeper integration.

Second, automated flags can be paired with playbooks and simple dashboards: for immediate support, trigger concrete interventions (e.g., sponsor a release manager for two cycles, offer contributor onboarding sprints, or adopt an RFC process to reduce PR latency). Recent OSS tools like OSSPREY \cite{khan2025ossprey}, offer actionable insights to sustainability issues, and can be readily extended with our findings.

\section{Threats To Validity}

We identify and address threats to validity across four dimensions: construct validity, external validity, internal validity, and reproducibility.

\subsection{Construct Validity}

\textbf{Sustainability vs. Success Operationalization.} A central construct validity concern is whether our operationalizations of ``sustainability'' (in foundations) and ``success'' (in GitHub) measure the same underlying phenomenon. As detailed in Section~\ref{sec:label_provenance} and Table~\ref{tab:label_provenance}, these labels are assigned by different authorities using different criteria: foundation sustainability requires demonstrated governance maturity, community diversity, IP compliance, and sustained releases, evaluated by PMCs or incubation committees over 1-3 year periods. GitHub success, as operationalized by Joblin et al.~\cite{6_joblin2022successful}, emphasizes contributor retention and activity momentum during early development phases, assigned retrospectively by researchers rather than institutional review boards.

Our RQ3 results empirically demonstrate that these are \textit{not} equivalent constructs: models trained on sustainability labels generalize reasonably to GitHub success (F1 = 85-93\%), but the reverse fails dramatically (F1 = 30-40\%). This asymmetry supports our theoretical position that sustainability subsumes certain success signals but is not reducible to them. However, this construct misalignment limits our ability to make strong causal claims about what factors drive sustainability universally vs.\ what factors drive success in informal OSS settings. Future work should develop unified sustainability measurement frameworks that can be consistently applied across governance contexts.

\textbf{Graduated as Sustainable, Retired as Unsustainable.} As noted in Section~2.2, project retirement does not always signify failure, a project may retire because its purpose has been fulfilled, superseded, or strategically de-prioritized at the foundation level. Nevertheless, treating graduation and retirement as the positive and negative classes, respectively, is a deliberate and well-grounded modeling choice: within the incubation window, the foundation's explicit goal is to evaluate whether a project has achieved self-sustaining community structure. Graduation operationalizes this outcome; retirement indicates that the project did not reach it within the evaluation period, regardless of the underlying cause. This is the standard operationalization in prior sustainability forecasting work~\cite{6_joblin2022successful}, and it provides a tractable, externally validated binary signal. We acknowledge that this framing introduces some label noise, particularly for strategically retired projects, but it reflects the most viable and reproducible definition available across foundations.

\textbf{Foundation Label Reliability.} Our sustainability labels depend on foundation-specific graduation/retirement decisions, which may reflect institutional biases, historical contingencies, or political factors beyond observable sociotechnical patterns. We partially mitigate this by focusing on the incubation period, when projects are actively evaluated, rather than post-graduation trajectories, but cannot fully eliminate institutional noise in ground-truth labels.

\subsection{External Validity}

\textbf{Generalization Beyond Incubator-Stage Projects.} Our analysis is explicitly restricted to projects during their incubation or early development phases. This design choice was intentional, to avoid label leakage and focus on decision-relevant prediction windows, but it limits generalization to mature, post-graduation project sustainability. The sociotechnical dynamics governing whether an established Top-Level Project (TLP) remains viable over decades may differ substantially from those predicting incubation outcomes. Our models should \textit{not} be applied to predict long-term sustainability of mature projects without retraining on post-graduation data.

\textbf{Survivorship Bias in Digital Trace Availability.} Our dataset is restricted to projects with sufficient digital traceability, commit histories, mailing list activity, and issue tracker records. Projects that left minimal digital footprints were excluded during preprocessing (Section~4.3), as sociotechnical network construction is not feasible without these traces. This introduces a form of survivorship bias: the models are trained on projects that were active enough to generate observable signals, which may not be representative of the full population of nascent OSS projects, including those that quietly stagnate with little recorded activity. This is an inherent constraint of any sociotechnical network-based approach, and one that is shared across the body of related work~\cite{6_joblin2022successful}. The practical implication is that our models are best suited for projects already exhibiting some activity, which aligns well with the target use case of foundation-stage sustainability triage.

\textbf{Temporal Confounds and Foundation Policy Evolution.} Our data collection spans projects from approximately 2005-2024, a period during which all three foundations evolved their governance processes, incubation criteria, and tooling. For example, ASF has iteratively refined its mentorship model and graduation requirements, while EF has adjusted its IP review procedures. These policy shifts introduce temporal confounds: sustainability patterns learned from earlier cohorts may not fully reflect what drives graduation in the current regime. We partially address this through temporal awareness in feature extraction and by restricting modeling to each project's incubation window rather than absolute calendar time. Periodic retraining on recent data remains necessary for production deployment, particularly as AI-assisted development, modern DevOps practices, and novel governance experiments (e.g., DAOs, token-based incentives) alter OSS collaboration norms.

\textbf{Representativeness of the Joblin et al. GitHub Dataset.} The GitHub component of our study relies on the curated dataset from Joblin et al.~\cite{6_joblin2022successful}, which classifies projects as successful or unsuccessful based on long-term contributor activity and community growth. This dataset represents a specific slice of the GitHub ecosystem, projects selected for analytical tractability, and may not be representative of GitHub OSS projects at large, which vary enormously in domain, size, ownership model, and lifecycle stage. We adopt this dataset because it provides the most carefully validated externally labeled GitHub sustainability proxy available in the literature, and because its labeling methodology (researcher-assigned, retrospective) is directly comparable to our incubation-based labels. The limitations of this dataset are documented in the original work and acknowledged as a constraint on the generalizability of our RQ3 findings.

\textbf{Foundation Coverage and Representativeness.} Our study covers three foundations (Apache, Eclipse, OSGeo) plus GitHub, representing substantial amount of major foundation governance archetypes based on our systematic analysis. However, we do not capture single-vendor controlled foundations (e.g., Android Open Source Project), cloud-native specialized foundations (CNCF), scientific computing foundations (NumFOCUS), or language-specific foundations (Python Software Foundation, Ruby Central). Additionally, our foundation sample is dominated by Western, English-language projects. The generalizability of our sociotechnical sustainability signals to non-English or regionally focused OSS governance remains an open question.

\subsection{Internal Validity}

\textbf{Class Imbalance.} Our datasets exhibit significant class imbalance, with graduated projects substantially outnumbering retired projects across all foundations (Apache: 205/57, Eclipse: 142/19, OSGeo: 13/7). This imbalance poses risks of model bias toward the majority class and unreliable minority class predictions. We addressed this through focal loss~\cite{111_mukhoti2020calibrating}, which explicitly down-weights easy examples and focuses learning on hard cases. The imbalance is particularly severe for OSGeo (7 retired projects) and GitHub (16 unsuccessful projects). We partially validate robustness through 5-fold stratified cross-validation and case study analysis of individual failed projects (Section~7), but acknowledge that predictions for retirement are less reliable than for graduation, especially in OF and GitHub contexts.

\textbf{Small Sample Sizes for OF and GitHub.} OSGeo (n=20 total) and GitHub (n=21 total) datasets are substantially smaller than Apache (n=262) and Eclipse (n=161). With only 7 retired OSGeo projects, a single misclassification in a 5-fold CV split represents 14\% error, making performance estimates noisy. The high performance of OSS-ProF on these datasets (F1 = 92.83\% for OF, 95.76\% for GitHub) should be understood as preliminary evidence requiring validation on larger samples rather than definitive proof of generalization. Future work should expand these datasets through additional data collection or explore data augmentation techniques (e.g., synthetic minority oversampling, time-series bootstrapping) to increase sample sizes while preserving temporal structure.

\textbf{Absence of Conventional ML Baselines.} We do not compare against conventional ML models (random forests, gradient boosting, logistic regression) because our data consists of variable-length temporal sequences (6-120+ months per project) with heterogeneous project scales. Conventional ML requires fixed-length feature vectors and does not natively handle sequential dependencies. While temporal aggregation or sliding window could enable baseline comparisons, these would sacrifice the evolutionary dynamics central to our approach. Our architecture choice follows prior work~\cite{6_joblin2022successful} demonstrating LSTMs effectiveness for temporal sustainability modeling.

\textbf{Model Instability Due to Limited Data.} The combination of small sample sizes and deep neural architectures introduces risk of model instability~\cite{52_kaplan2016note}. We mitigated this by running each experiment fifteen times and retaining median-performance model weights, ensuring reported results reflect typical rather than best-case performance. We provide all model weights as artifacts (see Section~\ref{sec:reproducibility}) to enable replication and sensitivity analysis.


\subsection{Reproducibility}
\label{sec:reproducibility}

\textbf{Artifact Availability.} To maximize reproducibility, we have made all code, data, and model weights publicly available through Zenodo (DOI: 10.5281/zenodo.14499305) and GitHub (\url{https://github.com/OSS-PREY/OSSPREY-Pex-Forecaster}). The artifact includes: (1) preprocessed sociotechnical network datasets for all 464 projects, (2) feature extraction and normalization scripts, (3) complete training code for all architectures (Bi-LSTM, Transformer, Bi-DLSTM), (4) trained model weights for all intra-foundation, cross-foundation, and OSS-ProF experiments, (5) SHAP-based interpretability analysis notebooks, and (6) detailed documentation of hyperparameters, random seeds, and software dependencies.

\textbf{Computational Environment.} Model training was conducted on a local server with specifications detailed in Appendix~\ref{local_server_config}, Table~\ref{server_config}. Our artifact includes a Docker container specification to minimize environment-related reproducibility issues.

\textbf{Data Preprocessing Transparency.} Our preprocessing pipeline (bot removal, de-aliasing, reply inference) involves multiple heuristic decisions that may affect downstream results. We have documented all preprocessing steps in detail (Section~4.2) and included the complete preprocessing codebase in our artifact. Sensitivity analysis of alternative preprocessing choices remains important future work.

\textbf{Stochastic Variation.} Despite setting random seeds, some variation in results may occur due to non-deterministic GPU operations in deep learning frameworks. Our reported metrics represent median performance across fifteen runs, but individual runs may deviate by $\pm$2-5\% in F1-score. Researchers attempting exact replication should expect results within this range.

\subsection{Mitigation Summary}

While we have identified substantive threats across all validity dimensions, we have taken explicit steps to mitigate their impact: (1) we clearly define and distinguish sustainability vs.\ success constructs and justify the graduated/retired label mapping; (2) we acknowledge survivorship bias as an inherent constraint of trace-based modeling and scope claims accordingly; (3) we address temporal confounds through incubation-window scoping and note the need for periodic retraining; (4) we document the provenance and limitations of the Joblin et al.\ GitHub dataset; (5) we address class imbalance through focal loss and validate on individual failed projects; and (6) we ensure full reproducibility through comprehensive artifact release. Future work should prioritize expanding foundation and platform coverage, increasing minority-class sample sizes, and developing unified sustainability measurement frameworks applicable across governance contexts.

\section{Conclusion}
This paper presents a comprehensive empirical study of sustainability prediction across three prominent open-source foundations, Apache (ASF), Eclipse (EF), and OSGeo (OF), as well as general GitHub projects. We used sociotechnical network features, trained deep learning models to forecast project sustainability, and examined their generalizability within and across ecosystems.

Our findings support the hypothesis that effective sustainability models can be developed within individual incubators. Cross-foundation training, particularly on combined ASFI and EFI datasets, further improves model robustness and predictive performance. However, we observed limitations in transferability: models trained on one incubator often degrade in performance when applied to another. This suggests that ecosystem-specific governance styles, project norms, and community engagement patterns shape unique sociotechnical dynamics that influence sustainability outcomes.

Extending our models to projects in GitHub but outside of foundations, we found that incubator-trained classifiers can retain predictive power. This is further evidence of the generalizability of the framework to general OSS projects. Our best performing approach, OSS-ProF, uses such profiles to match projects to foundation models, with effective performance on sustainability prediction.

Our analysis of feature importance reveals that sustainability is influenced by both technical and social engagement, with the relative contribution of each varying across foundations. 

To our knowledge, this is the first study to systematically investigate sustainability prediction across multiple OSS incubators and in GitHub (outside of foundations). Our work demonstrates that sociotechnical networks offer a powerful abstraction for sustainability modeling and that careful attention to ecosystem-specific signals is essential for developing  models that can be used in practice. Future research should explore hybrid architectures, contrastive learning, and interpretable modeling to further improve generalization and practical utility.

\section{Data \& Code Availability Statement}
The source code and data of this research are available at: \href{https://zenodo.org/records/18676576}{this link (https://zenodo.org/records/18676576)}.

\section{Declaration of Interests}
The authors declare that they have no known competing financial interests or personal relationships that could have appeared to influence the work reported in this paper.

\section{Acknowledgments}
The research received funding from the National Science Foundation under Grant No. 2020751


\bibliographystyle{unsrt}
\bibliography{references}

\begin{thebibliography}{99}

\bibitem{1_yin2023self}
Yin, Likang and Zhang, Xiyu and Filkov, Vladimir. On the Self-Governance and Episodic Changes in Apache Incubator Projects: An Empirical Study. In 2023 IEEE/ACM 45th International Conference on Software Engineering (ICSE). pp. 678--689. 2023.

\bibitem{2_yin2021sustainability}
Yin, Likang and Chen, Zhuangzhi and Xuan, Qi and Filkov, Vladimir. Sustainability forecasting for Apache Incubator projects. In Proceedings of the 29th ACM joint meeting on European Software Engineering Conference and Symposium on the Foundations of Software Engineering. pp. 1056--1067. 2021.

\bibitem{3_yin2021apache}
Yin, Likang and Zhang, Zhiyuan and Xuan, Qi and Filkov, Vladimir. Apache software foundation incubator project sustainability dataset. In 2021 ieee/acm 18th international conference on mining software repositories (msr). pp. 595--599. 2021.

\bibitem{4_xiao2023early}
Xiao, Wenxin and He, Hao and Xu, Weiwei and Zhang, Yuxia and Zhou, Minghui. How early participation determines long-term sustained activity in github projects?. In Proceedings of the 31st ACM Joint European Software Engineering Conference and Symposium on the Foundations of Software Engineering. pp. 29--41. 2023.

\bibitem{5_ait2022empirical}
Ait, Adem and Izquierdo, Javier Luis C{\'a}novas and Cabot, Jordi. An empirical study on the survival rate of GitHub projects. In Proceedings of the 19th International Conference on Mining Software Repositories. pp. 365--375. 2022.

\bibitem{6_joblin2022successful}
Joblin, Mitchell and Apel, Sven. How do successful and failed projects differ? a socio-technical analysis. ACM Transactions on Software Engineering and Methodology (TOSEM) 31(4):1--24. 2022.

\bibitem{7_wessel2018power}
Wessel, Mairieli and De Souza, Bruno Mendes and Steinmacher, Igor and Wiese, Igor S and Polato, Ivanilton and Chaves, Ana Paula and Gerosa, Marco A. The power of bots: Characterizing and understanding bots in oss projects. Proceedings of the ACM on Human-Computer Interaction 2(CSCW):1--19. 2018.

\bibitem{8_wessel2021don}
Wessel, Mairieli and Wiese, Igor and Steinmacher, Igor and Gerosa, Marco Aurelio. Don't disturb me: Challenges of interacting with software bots on open source software projects. Proceedings of the ACM on Human-Computer Interaction 5(CSCW2):1--21. 2021.

\bibitem{9_vasilescu2015gender}
Vasilescu, Bogdan and Posnett, Daryl and Ray, Baishakhi and van den Brand, Mark GJ and Serebrenik, Alexander and Devanbu, Premkumar and Filkov, Vladimir. Gender and tenure diversity in GitHub teams. In Proceedings of the 33rd annual ACM conference on human factors in computing systems. pp. 3789--3798. 2015.

\bibitem{10_1dilated}
Chang, Shiyu and Zhang, Yang and Han, Wei and Yu, Mo and Guo, Xiaoxiao and Tan, Wei and Cui, Xiaodong and Witbrock, Michael and Hasegawa-Johnson, Mark A and Huang, Thomas S. Dilated recurrent neural networks. Advances in neural information processing systems 30. 2017.

\bibitem{10_2dilated}
Wang, Rongxi and Peng, Caiyuan and Gao, Jianmin and Gao, Zhiyong and Jiang, Hongquan. A dilated convolution network-based LSTM model for multi-step prediction of chaotic time series. Computational and Applied Mathematics 39:1--22. 2020.

\bibitem{10_3dilated}
Borovykh, Anastasia and Bohte, Sander and Oosterlee, Cornelis W. Dilated convolutional neural networks for time series forecasting. Journal of Computational Finance, Forthcoming. 2018.

\bibitem{11_GRU}
Zhou, Qihang and Zhou, Changjun and Wang, Xiao. Stock prediction based on bidirectional gated recurrent unit with convolutional neural network and feature selection. PloS one 17(2):e0262501. 2022.

\bibitem{12_BiLSTM}
Khan, Mehak and Wang, Hongzhi and Riaz, Adnan and Elfatyany, Aya and Karim, Sajida. Bidirectional LSTM-RNN-based hybrid deep learning frameworks for univariate time series classification. The Journal of Supercomputing 77:7021--7045. 2021.

\bibitem{13_khan2024models}
Khan, Nafiz Imtiaz and Filkov, Vladimir. From Models to Practice: Enhancing OSS Project Sustainability with Evidence-Based Advice. In Companion Proceedings of the 32nd ACM International Conference on the Foundations of Software Engineering. pp. 457–461. 2024. doi: \url{https://doi.org/10.1145/3663529.3663777}. \url{https://doi.org/10.1145/3663529.3663777}.

\bibitem{14_hutter2019automated}
Hutter, Frank and Kotthoff, Lars and Vanschoren, Joaquin. Automated machine learning: methods, systems, challenges. Springer Nature. 2019.

\bibitem{15_1apachefoundation}
Yang, Nan and Ferreira, Isabella and Serebrenik, Alexander and Adams, Bram. Why do projects join the apache software foundation?. In Proceedings of the 2022 ACM/IEEE 44th international conference on software engineering: software engineering in society. pp. 161--171. 2022.

\bibitem{15_2apachefoundation}
Eilebrecht, Lars. Behind the Scenes of The Apache Software Foundation. In Software libre: II Conferencia Internacional. pp. 168--175. 2006.

\bibitem{15_3gharehyazie2015developer}
Gharehyazie, Mohammad and Posnett, Daryl and Vasilescu, Bogdan and Filkov, Vladimir. Developer initiation and social interactions in OSS: A case study of the Apache Software Foundation. Empirical Software Engineering 20:1318--1353. 2015.

\bibitem{17_1eclipsefoundation}
Aarnoutse, Floor and Renes, Cassandra and Snijders, Remco and Jansen, Slinger. The reality of an associate model: Comparing partner activity in the eclipse ecosystem. In Proceedings of the 2014 European Conference on Software Architecture Workshops. pp. 1--6. 2014.

\bibitem{17_2eclipsefoundation}
Mizushima, Kazunori and Ikawa, Yasuo. A structure of co-creation in an open source software ecosystem: A case study of the eclipse community. In 2011 Proceedings of PICMET'11: Technology Management in the Energy Smart World (PICMET). pp. 1--8. 2011.

\bibitem{17_3eclipsefoundation}
Frost, Randall. Jazz and the eclipse way of collaboration. IEEE software 24(6):114--117. 2007.

\bibitem{18_github_distribution}
McDonald, Nora and Blincoe, Kelly and Petakovic, EVA and Goggins, Sean. Modeling distributed collaboration on Github. Advances in Complex Systems 17(07n08):1450024. 2014.

\bibitem{19_github_code_of_conduct}
Li, Renee and Pandurangan, Pavitthra and Frluckaj, Hana and Dabbish, Laura. Code of conduct conversations in open source software projects on github. Proceedings of the ACM on Human-computer Interaction 5(CSCW1):1--31. 2021.

\bibitem{20_ait2022empirical}
Ait, Adem and Izquierdo, Javier Luis C{\'a}novas and Cabot, Jordi. An empirical study on the survival rate of GitHub projects. In Proceedings of the 19th International Conference on Mining Software Repositories. pp. 365--375. 2022.

\bibitem{21_mazziotta2022normalization}
Mazziotta, Matteo and Pareto, Adriano. Normalization methods for spatio-temporal analysis of environmental performance: Revisiting the Min--Max method. Environmetrics 33(5):e2730. 2022.

\bibitem{22_fei2021z}
Fei, Nanyi and Gao, Yizhao and Lu, Zhiwu and Xiang, Tao. Z-score normalization, hubness, and few-shot learning. In Proceedings of the IEEE/CVF International Conference on Computer Vision. pp. 142--151. 2021.

\bibitem{23_apacheway}
Curcuru, Shane. If it didn’t happen on the mailing list, it didn’t happen. 2018. \url{https://theapacheway.com/on-list/}. Accessed: 2024-07-09.

\bibitem{24_eclipse_incubation}
Eclipse Foundation. What is Incubation?. 2024. \url{https://wiki.eclipse.org/Development_Resources/Process_Guidelines/What_is_Incubation}. Accessed: 2024-07-09.

\bibitem{25_rahman2014insight}
Rahman, Mohammad Masudur and Roy, Chanchal K. An insight into the pull requests of github. In Proceedings of the 11th working conference on mining software repositories. pp. 364--367. 2014.

\bibitem{26_ren2020starin}
Ren, Leiming and Shan, Shimin and Xu, Xiujuan and Liu, Yu. Starin: An approach to predict the popularity of github repository. In Data Science: 6th International Conference of Pioneering Computer Scientists, Engineers and Educators, ICPCSEE 2020, Taiyuan, China, September 18-21, 2020, Proceedings, Part II 6. pp. 258--273. 2020.

\bibitem{27_coelho2018identifying}
Coelho, Jailton and Valente, Marco Tulio and Silva, Luciana L and Shihab, Emad. Identifying unmaintained projects in github. In Proceedings of the 12th ACM/IEEE International Symposium on Empirical Software Engineering and Measurement. pp. 1--10. 2018.

\bibitem{28_comino2007planning}
Comino, Stefano and Manenti, Fabio M and Parisi, Maria Laura. From planning to mature: On the success of open source projects. Research policy 36(10):1575--1586. 2007.

\bibitem{29_weber2004success}
Weber, Steven. The success of open source. Harvard University Press. 2004.

\bibitem{30_crowston2003defining}
Crowston, Kevin and Annabi, Hala and Howison, James. Defining open source software project success. 2003.

\bibitem{31_gamalielsson2014sustainability}
Gamalielsson, Jonas and Lundell, Bj{\"o}rn. Sustainability of Open Source software communities beyond a fork: How and why has the LibreOffice project evolved?. Journal of systems and Software 89:128--145. 2014.

\bibitem{32_wu2007investigating}
Wu, Jing and Goh, Khim-Yong and Tang, Qian. Investigating success of open source software projects: A social network perspective. ICIS 2007 Proceedings:105. 2007.

\bibitem{33_wu2007analysis}
Wu, Jing and Tang, Qian. Analysis of survival of open source projects: A social network perspective. PACIS 2007 Proceedings:19. 2007.

\bibitem{34_van2022tractability}
Van den Broeck, Guy and Lykov, Anton and Schleich, Maximilian and Suciu, Dan. On the tractability of SHAP explanations. Journal of Artificial Intelligence Research 74:851--886. 2022.

\bibitem{35_fitzgerald2006transformation}
Fitzgerald, Brian. The transformation of open source software. MIS quarterly:587--598. 2006.

\bibitem{36_hawkins2004problem}
Hawkins, Douglas M. The problem of overfitting. Journal of chemical information and computer sciences 44(1):1--12. 2004.

\bibitem{37_grid}
Syarif, Iwan and Prugel-Bennett, Adam and Wills, Gary. SVM parameter optimization using grid search and genetic algorithm to improve classification performance. TELKOMNIKA (Telecommunication Computing Electronics and Control) 14(4):1502--1509. 2016.

\bibitem{38_residual}
Luo, Jian-Hao and Wu, Jianxin. Neural network pruning with residual-connections and limited-data. In Proceedings of the IEEE/CVF conference on computer vision and pattern recognition. pp. 1458--1467. 2020.

\bibitem{39_goutte2005probabilistic}
Goutte, Cyril and Gaussier, Eric. A probabilistic interpretation of precision, recall and F-score, with implication for evaluation. In European conference on information retrieval. pp. 345--359. 2005.

\bibitem{40_gitnux2024}
{Gitnux}. Open Source Software Statistics. 2024. \url{https://gitnux.org/open-source-software-statistics/}.

\bibitem{41_schweik2012internet}
Schweik, Charles M and English, Robert C. Internet success: a study of open-source software commons. MIT Press. 2012.

\bibitem{42_clements2016real}
Clements, Michael P. Real-time factor model forecasting and the effects of instability. Computational Statistics \& Data Analysis 100:661--675. 2016.

\bibitem{43_park2024survivability}
Park, Sohee and Kwon, Ryeonggu and Kwon, Gihwon. Survivability Prediction of Open Source Software with Polynomial Regression. Applied Sciences 14(7):2812. 2024.

\bibitem{44_niu2021review}
Niu, Zhaoyang and Zhong, Guoqiang and Yu, Hui. A review on the attention mechanism of deep learning. Neurocomputing 452:48--62. 2021.

\bibitem{45_goodfellow2016deep}
Goodfellow, Ian and Bengio, Yoshua and Courville, Aaron. Deep learning. MIT press. 2016.

\bibitem{46_sosna2010business}
Sosna, Marc and Trevinyo-Rodr{\'\i}guez, Rosa Nelly and Velamuri, S Ramakrishna. Business model innovation through trial-and-error learning: The Naturhouse case. Long range planning 43(2-3):383--407. 2010.

\bibitem{47_laeddine2021deep}
Alaeddine, Hmidi and Jihene, Malek. Deep residual network in network. Computational Intelligence and Neuroscience 2021(1):6659083. 2021.

\bibitem{48_prechelt2002early}
Prechelt, Lutz. Early stopping-but when?. In Neural Networks: Tricks of the trade. pp. 55--69. 2002.

\bibitem{49_jethani2021fastshap}
Jethani, Neil and Sudarshan, Mukund and Covert, Ian Connick and Lee, Su-In and Ranganath, Rajesh. Fastshap: Real-time shapley value estimation. In International conference on learning representations. 2021.

\bibitem{50_takahashi2022confidence}
Takahashi, Kanae and Yamamoto, Kouji and Kuchiba, Aya and Koyama, Tatsuki. Confidence interval for micro-averaged F 1 and macro-averaged F 1 scores. Applied Intelligence 52(5):4961--4972. 2022.

\bibitem{51_gu2022complex}
Gu, Zuguang. Complex heatmap visualization. Imeta 1(3):e43. 2022.

\bibitem{52_kaplan2016note}
Kaplan, Andee and Nordman, Daniel and Vardeman, Stephen. A note on the instability and degeneracy of deep learning models. arXiv preprint arXiv:1612.01159. 2016.

\bibitem{53_japkowicz2002class}
Japkowicz, Nathalie and Stephen, Shaju. The class imbalance problem: A systematic study. Intelligent data analysis 6(5):429--449. 2002.

\bibitem{54_drummond2003c4}
Drummond, Chris and Holte, Robert C and others. C4. 5, class imbalance, and cost sensitivity: why under-sampling beats over-sampling. In Workshop on learning from imbalanced datasets II. 2003.

\bibitem{55_artstein2017inter}
Artstein, Ron. Inter-annotator agreement. Handbook of linguistic annotation:297--313. 2017.

\bibitem{60_ter1986weighted}
Ter Braak, Cajo JF and Looman, Caspar WN. Weighted averaging, logistic regression and the Gaussian response model. Vegetatio 65:3--11. 1986.

\bibitem{61_kakao}
KakaoTalk. 2024. \url{https://www.kakaocorp.com/page/service/service/KakaoTalk?lang=en}. Accessed: 2024-08-01.

\bibitem{62_apache}
Apache Software Foundation. 2024. \url{https://www.apache.org/}. Accessed: 2024-08-01.

\bibitem{eclipse}
Eclipse Foundation. 2024. \url{https://www.eclipse.org/org/}. Accessed: 2024-08-01.

\bibitem{github}
GitHub. 2024. \url{https://github.com/}. Accessed: 2024-08-01.

\bibitem{gist}
Gist. 2024. \url{https://gist.github.com/ppisarczyk/43962d06686722d26d176fad46879d41}. Accessed: 2024-08-03.

\bibitem{OSSPREY_OSS_Scraper_Tool}
Khan, Nafiz Imtiaz and Soni, Priyanshu and Ashok, Abhishek and Kashyap, Sankalp and Filkov, Vladimir. OSSPREY-OSS-Scraper-Tool: Open Source Sustainability Scraper for GitHub Projects. 2025. \url{https://github.com/OSS-PREY/OSSPREY-OSS-Scraper-Tool}. GitHub repository.

\bibitem{rust}
Rust Programming Language. 2024. \url{https://www.rust-lang.org/}. Accessed: 2024-08-03.

\bibitem{linuxfoundation}
The Linux Foundation. 2024. \url{https://www.linuxfoundation.org/}. Accessed: 2024-08-03.

\bibitem{osgeo}
{Open Source Geospatial Foundation}. OSGeo – The Open Source Geospatial Foundation. 2025. \url{https://www.osgeo.org/}. Accessed: 2025-06-03.

\bibitem{genomefoundation}
The Soccer Genome Foundation. 2024. \url{https://www.thesoccergenomefoundation.org/}. Accessed: 2024-08-03.

\bibitem{python}
Python Programming Language. 2024. \url{https://www.python.org/}. Accessed: 2024-08-03.

\bibitem{agarap2018deep}
Agarap, Abien Fred. Deep learning using rectified linear units (relu). arXiv preprint arXiv:1803.08375. 2018.

\bibitem{gczhang2019gradient}
Zhang, Jingzhao and He, Tianxing and Sra, Suvrit and Jadbabaie, Ali. Why gradient clipping accelerates training: A theoretical justification for adaptivity. arXiv preprint arXiv:1905.11881. 2019.

\bibitem{100_ramchandran2022exploring}
Ramchandran, Anirudh and Yin, Likang and Filkov, Vladimir. Exploring Apache incubator project trajectories with APEX. In Proceedings of the 19th international conference on mining software repositories. pp. 333--337. 2022.

\bibitem{101_yin2022open}
Yin, Likang and Chakraborti, Mahasweta and Yan, Yibo and Schweik, Charles and Frey, Seth and Filkov, Vladimir. Open source software sustainability: Combining institutional analysis and socio-technical networks. Proceedings of the ACM on Human-Computer Interaction 6(CSCW2):1--23. 2022.

\bibitem{102_sarker2019socio}
Sarker, Farhana and Vasilescu, Bogdan and Blincoe, Kelly and Filkov, Vladimir. Socio-technical work-rate increase associates with changes in work patterns in online projects. In 2019 IEEE/ACM 41st International Conference on Software Engineering (ICSE). pp. 936--947. 2019.

\bibitem{103_bird2009putting}
Bird, Christian and Nagappan, Nachiappan and Gall, Harald and Murphy, Brendan and Devanbu, Premkumar. Putting it all together: Using socio-technical networks to predict failures. In 2009 20th International Symposium on Software Reliability Engineering. pp. 109--119. 2009.

\bibitem{104_appelbaum1997socio}
Appelbaum, Steven H. Socio-technical systems theory: an intervention strategy for organizational development. Management decision 35(6):452--463. 1997.

\bibitem{105_hong2017creating}
Hong, Jinwon and Lee, One-Ki and Suh, Woojong. Creating knowledge within a team: a socio-technical interaction perspective. Knowledge management research \& practice 15(1):23--33. 2017.

\bibitem{106_howison2011validity}
Howison, James and Wiggins, Andrea and Crowston, Kevin. Validity issues in the use of social network analysis with digital trace data. Journal of the Association for Information Systems 12(12):2. 2011.

\bibitem{107_scacchi2005socio}
Scacchi, Walt. Socio-technical interaction networks in free/open source software development processes. In Software process modeling. pp. 1--27. 2005.

\bibitem{108_storey2020software}
Storey, Margaret-Anne and Ernst, Neil A and Williams, Courtney and Kalliamvakou, Eirini. The who, what, how of software engineering research: a socio-technical framework. Empirical Software Engineering 25:4097--4129. 2020.

\bibitem{109_philipp2017exploding}
Philipp, George and Song, Dawn and Carbonell, Jaime G. The exploding gradient problem demystified-definition, prevalence, impact, origin, tradeoffs, and solutions. arXiv preprint arXiv:1712.05577. 2017.

\bibitem{vaswani2017attention}
Vaswani, Ashish and Shazeer, Noam and Parmar, Niki and Uszkoreit, Jakob and Jones, Llion and Gomez, Aidan N and Kaiser, {\L}ukasz and Polosukhin, Illia. Attention is all you need. Advances in neural information processing systems 30. 2017.

\bibitem{110_rozinek2024fast}
Rozinek, Ond{\v{r}}ej and Mare{\v{s}}, Jan. Fast and precise convolutional jaro and jaro-Winkler similarity. In 2024 35th Conference of Open Innovations Association (FRUCT). pp. 604--613. 2024.

\bibitem{111_mukhoti2020calibrating}
Mukhoti, Jishnu and Kulharia, Viveka and Sanyal, Amartya and Golodetz, Stuart and Torr, Philip and Dokania, Puneet. Calibrating deep neural networks using focal loss. Advances in neural information processing systems 33:15288--15299. 2020.

\bibitem{joblin2022successful}
Joblin, Mitchell and Apel, Sven. How do successful and failed projects differ? A socio-technical analysis. ACM Transactions on Software Engineering and Methodology (TOSEM) 31(4):1--24. 2022.

\bibitem{khan2025ossprey}
Khan, Nafiz I and Soni, Priyal and Ashok, Arjun and Filkov, Vladimir. OSSPREY: AI-Driven Forecasting and Intervention for OSS Project Sustainability. ASE 2025. 2025.

\bibitem{mockus2007large}
Mockus, Audris. Large-scale code reuse in open source software. In First International Workshop on Emerging Trends in FLOSS Research and Development (FLOSS'07: ICSE Workshops 2007). pp. 7--7. 2007.

\bibitem{jergensen2011onion}
Jergensen, Corey and Sarma, Anita and Wagstrom, Patrick. The onion patch: migration in open source ecosystems. In Proceedings of the 19th ACM SIGSOFT symposium and the 13th European conference on Foundations of software engineering. pp. 70--80. 2011.

\bibitem{gaughan2009examination}
Gaughan, Gary and Fitzgerald, Brian and Shaikh, Maha. An examination of the use of open source software processes as a global software development solution for commercial software engineering. In 2009 35th Euromicro Conference on Software Engineering and Advanced Applications. pp. 20--27. 2009.

\bibitem{steiniger20132012}
Steiniger, Stefan and Hunter, Andrew JS. The 2012 free and open source GIS software map--A guide to facilitate research, development, and adoption. Computers, environment and urban systems 39:136--150. 2013.

\bibitem{ramseystate}
Ramsey, Paul. The State of Open Source GIS.

\bibitem{contingency_ref_1}
Lawrence, Paul R. and Lorsch, Jay W. Organization and Environment: Managing Differentiation and Integration. Harvard Business School Press. 1967.

\bibitem{contingency_ref_2}
Galbraith, Jay R. Designing Complex Organizations. Addison-Wesley Longman Publishing Co., Inc. 1973.

\bibitem{GitHub}
GitHub. \emph{GitHub}. \url{https://github.com}.

\bibitem{18_GitHub_distribution}
GitHub. \emph{GitHub: About the GitHub community and usage (documentation)}. \url{https://docs.github.com}.

\bibitem{19_GitHub_code_of_conduct}
GitHub. \emph{GitHub Community Code of Conduct}. \url{https://docs.github.com/en/site-policy/github-terms/github-community-code-of-conduct}.

\end{thebibliography}

\newpage

\clearpage

\begin{table*}[!ht]
\section{Local Server Configuration}
\label{local_server_config}
\centering
\caption{Local Server Configuration}
\label{server_config}
{
\begin{tabular}{ll}
\hline
\textbf{Component} & \textbf{Configuration} \\
\hline
Primary Memory (RAM) & 64 GB \\
\hline
Secondary Memory (Hard Disk) & 512 GB SSD \\
\hline
Processor & 
\begin{tabular}{@{}l@{}}
\textit{Name}: Intel(R) Xeon(R) W-2135 CPU @ 3.70GHz \\
\textit{Architecture}: x86\_64 \\
\textit{CPU(s)}: 12 \\
\textit{Thread(s) per core}: 2 \\
\textit{Core(s) per socket}: 6 \\
\textit{Max Clock Speed}: 4500.00 MHz \\
\textit{Min Clock Speed}: 1200.00 MHz \\
\textit{PCI Express Lanes}: 48 \\
\textit{Integrated Graphics}: None \\
\textit{TDP (Thermal Design Power)}: 140 W \\
\end{tabular} \\
\hline
Cache & 
\begin{tabular}{@{}l@{}}
\textit{L1d cache}: 192 KiB (6 instances) \\
\textit{L1i cache}: 192 KiB (6 instances) \\
\textit{L2 cache}: 6 MiB (6 instances) \\
\textit{L3 cache}: 8.3 MiB (1 instance) \\
\end{tabular} \\
\hline
GPU & 
\begin{tabular}{@{}l@{}}
\textit{Name}: GV100 [TITAN V] \\
\textit{Memory}: 12 GB HBM2 \\
\textit{Memory Interface Width}: 3072 bits \\
\textit{Memory Bandwidth}: 652.8 GB/s \\
\textit{Base Clock Speed}: 1200 MHz \\
\textit{Boost Clock Speed}: 1455 MHz \\
\textit{Memory Clock Speed}: 850 MHz (1700 MHz effective) \\
\textit{CUDA Cores}: 5120 \\
\textit{Tensor Cores}: 640 \\
\textit{Thermal Design Power (TDP)}: 250 W \\
\end{tabular} \\
\hline
\end{tabular}

}
\end{table*}


\begin{table*}[!ht]
\section{Cross Validation Variance}
\label{cv_variance}
\centering
\caption{Standard deviation of each trial's results for the best performing model.}
\label{tab:std_dev_results}
\renewcommand{\arraystretch}{0.75}
\begin{tabular}{lrr}
\toprule
Strategy & Std Dev & F1-Score \\
\midrule
A $\to$ E & 0.0315 & \textbf{0.7176} \\
A $\to$ G & 0.0264 & \textbf{0.9505} \\
A $\to$ O & 0.0000 & \textbf{0.7558} \\
E $\to$ E & 0.0134 & \textbf{0.9063} \\
G $\to$ E & 0.0113 & \textbf{0.5639} \\
G $\to$ O & 0.0036 & \textbf{0.7063} \\
E $\to$ A & 0.0209 & \textbf{0.7487} \\
A $\to$ A & 0.0095 & \textbf{0.9615} \\
E $\to$ G & 0.0235 & \textbf{0.8095} \\
E $\to$ O & 0.0626 & \textbf{0.7288} \\
G $\to$ A & 0.0692 & \textbf{0.7774} \\
O $\to$ O & 0.1242 & \textbf{0.9286} \\
O $\to$ G & 0.0284 & \textbf{0.8613} \\
O $\to$ E & 0.0455 & \textbf{0.8937} \\
O $\to$ A & 0.0297 & \textbf{0.8480} \\
G $\to$ G & 0.0583 & \textbf{1.0000} \\
A + O $\to$ A & 0.0146 & \textbf{0.9454} \\
E + O $\to$ A & 0.0199 & \textbf{0.7231} \\
E $\to$ A + O & 0.0208 & \textbf{0.7305} \\
E + O $\to$ G & 0.0129 & \textbf{0.7913} \\
E + O $\to$ E & 0.0044 & \textbf{0.8938} \\
A $\to$ E + O & 0.0202 & \textbf{0.6993} \\
A + E $\to$ A & 0.0130 & \textbf{0.9231} \\
A + E $\to$ O & 0.1077 & \textbf{0.7558} \\
O + A $\to$ O & 0.0192 & \textbf{0.7667} \\
O + E $\to$ O & 0.1347 & \textbf{0.7667} \\
O $\to$ A + E & 0.0125 & \textbf{0.8120} \\
A + E $\to$ G & 0.1367 & \textbf{0.8957} \\
A + O $\to$ E & 0.0156 & \textbf{0.6820} \\
A + O $\to$ G & 0.0264 & \textbf{0.9505} \\
A + E $\to$ E & 0.0311 & \textbf{0.7681} \\
A + E + O $\to$ G & 0.0080 & \textbf{0.9095} \\
O + A + E $\to$ O & 0.0997 & \textbf{0.7667} \\
A + E $\to$ A + E & 0.0074 & \textbf{0.8702} \\
A + O $\to$ A + O & 0.0087 & \textbf{0.9264} \\
E + O $\to$ E + O & 0.0345 & \textbf{0.8573} \\
A + E + O $\to$ A & 0.0221 & \textbf{0.9231} \\
A + E + O $\to$ E & 0.0296 & \textbf{0.7587} \\
A + E $\to$ A + E + G & 0.0131 & \textbf{0.8810} \\
A + E + O $\to$ A + E + O & 0.0136 & \textbf{0.8668} \\
A + E + O $\to$ A + E + O + G & 0.0181 & \textbf{0.8654} \\
\bottomrule
\end{tabular}
\end{table*}

\begin{table}[!ht]
\section{Foundation Coverage}
\label{foundation_coverage}
\vspace{-0.75cm}
\centering
\caption{Foundation Characteristics Coverage Matrix}
\label{tab:foundation_coverage}

\resizebox{\textwidth}{!}{%
\setlength{\arrayrulewidth}{0.7pt} 
\renewcommand{\arraystretch}{0.55} 
\begin{tabular}{@{}l|ccc|l@{}}
\toprule
\textbf{Characteristic} & \textbf{Apache} & \textbf{Eclipse} & \textbf{OSGeo} & \textbf{Status} \\
\cmidrule(lr){2-4} 
\midrule

\multicolumn{5}{@{}l}{\textbf{\textit{Governance Models}}} \\
\specialrule{0.8pt}{0.25em}{0.25em}
\rowcolor{gray!6} Meritocratic/Community-driven & \checkmark & & & \cellcolor{green!20}Covered \\
Corporate-backed & & \checkmark & & \cellcolor{green!20}Covered \\
\rowcolor{gray!6} Vendor-neutral governance & & \checkmark & & \cellcolor{green!20}Covered \\
Domain-specific expert review & & & \checkmark & \cellcolor{green!20}Covered \\
\rowcolor{gray!6} Single-vendor control & & & & \cellcolor{red!20}Not Covered \\
Benevolent dictator model & & & & \cellcolor{red!20}Not Covered \\
\specialrule{0.8pt}{0.35em}{0.35em}

\multicolumn{5}{@{}l}{\textbf{\textit{Decision-Making Processes}}} \\
\specialrule{0.8pt}{0.25em}{0.25em}
\rowcolor{gray!6} Consensus voting & \checkmark & & & \cellcolor{green!20}Covered \\
PMC oversight & & \checkmark & & \cellcolor{green!20}Covered \\
\rowcolor{gray!6} Multi-tier expert review & & & \checkmark & \cellcolor{green!20}Covered \\
Technical steering committees & & & & \cellcolor{red!20}Not Covered \\
\specialrule{0.8pt}{0.35em}{0.35em}

\multicolumn{5}{@{}l}{\textbf{\textit{Communication \& Collaboration}}} \\
\specialrule{0.8pt}{0.25em}{0.25em}
\rowcolor{gray!6} Mandatory mailing lists & \checkmark & & & \cellcolor{green!20}Covered \\
Flexible communication channels & & \checkmark & \checkmark & \cellcolor{green!20}Covered \\
\rowcolor{gray!6} Conference-driven engagement & & & \checkmark & \cellcolor{green!20}Covered \\
Real-time chat platforms (Slack/Discord) & & & & \cellcolor{red!20}Not Covered \\
\specialrule{0.8pt}{0.35em}{0.35em}

\multicolumn{5}{@{}l}{\textbf{\textit{Incubation \& Project Lifecycle}}} \\
\specialrule{0.8pt}{0.25em}{0.25em}
\rowcolor{gray!6} Strict process adherence (Apache Way) & \checkmark & & & \cellcolor{green!20}Covered \\
IP-focused legal review & & \checkmark & & \cellcolor{green!20}Covered \\
\rowcolor{gray!6} Flexible multi-tier pathways & & & \checkmark & \cellcolor{green!20}Covered \\
Sandbox/experimental tiers & & & & \cellcolor{red!20}Not Covered \\
\specialrule{0.8pt}{0.35em}{0.35em}

\multicolumn{5}{@{}l}{\textbf{\textit{Funding Models}}} \\
\specialrule{0.8pt}{0.25em}{0.25em}
\rowcolor{gray!6} Donation-based & \checkmark & & & \cellcolor{green!20}Covered \\
Corporate membership fees & & \checkmark & & \cellcolor{green!20}Covered \\
\rowcolor{gray!6} Grant + conference revenue & & & \checkmark & \cellcolor{green!20}Covered \\
Sponsorship/patronage model & & & & \cellcolor{red!20}Not Covered \\
\specialrule{0.8pt}{0.35em}{0.35em}

\multicolumn{5}{@{}l}{\textbf{\textit{Community Composition}}} \\
\specialrule{0.8pt}{0.25em}{0.25em}
\rowcolor{gray!6} Volunteer-driven communities & \checkmark & & & \cellcolor{green!20}Covered \\
Corporate contributors & & \checkmark & & \cellcolor{green!20}Covered \\
\rowcolor{gray!6} Academic/research communities & & & \checkmark & \cellcolor{green!20}Covered \\
Hobbyist/individual developers & & & & \cellcolor{red!20}Not Covered \\
\specialrule{0.8pt}{0.35em}{0.35em}

\multicolumn{5}{@{}l}{\textbf{\textit{Project Scale \& Domain Focus}}} \\
\specialrule{0.8pt}{0.25em}{0.25em}
\rowcolor{gray!6} Large general-purpose (350+ projects) & \checkmark & & & \cellcolor{green!20}Covered \\
Large enterprise-focused (420+ projects) & & \checkmark & & \cellcolor{green!20}Covered \\
\rowcolor{gray!6} Small domain-specific (68 projects) & & & \checkmark & \cellcolor{green!20}Covered \\
Infrastructure/kernel focus & & & & \cellcolor{red!20}Not Covered \\
\rowcolor{gray!6} Cloud-native specialization & & & & \cellcolor{red!20}Not Covered \\
Language-specific focus & & & & \cellcolor{red!20}Not Covered \\
\rowcolor{gray!6} Scientific computing focus & & & & \cellcolor{red!20}Not Covered \\
\specialrule{0.8pt}{0.35em}{0.35em}

\multicolumn{5}{@{}l}{\textbf{\textit{Licensing \& Intellectual Property}}} \\
\specialrule{0.8pt}{0.25em}{0.25em}
\rowcolor{gray!6} Apache License emphasis & \checkmark & & & \cellcolor{green!20}Covered \\
Rigorous IP provenance tracking & & \checkmark & & \cellcolor{green!20}Covered \\
\rowcolor{gray!6} Flexible licensing approaches & & & \checkmark & \cellcolor{green!20}Covered \\
Dual licensing strategies & & & & \cellcolor{red!20}Not Covered \\
\rowcolor{gray!6} Mixed proprietary/OSS models & & & & \cellcolor{red!20}Not Covered \\
\bottomrule
\end{tabular}%
}
\end{table}


\end{document}